\documentclass[a4paper,aps,pre,superscriptaddress,floatfix,nofootinbib,longbibliography,notitlepage,twocolumn]{revtex4-1}
\usepackage{bbm}
\usepackage{bbold}
\usepackage{bbm}
\usepackage[pdftex]{graphicx}
\usepackage{latexsym,amsmath,verbatim,amssymb,txfonts}
\usepackage{lipsum}
\usepackage{hyperref}
\usepackage{color}
\usepackage{svg}
\usepackage{rotating}
\usepackage{verbatim}
\usepackage{multirow}
\usepackage[english]{babel}
\usepackage{comment}
\usepackage{bm}
\usepackage{amsmath}
\usepackage{amsfonts}
\usepackage{amssymb}
\usepackage{color}
\usepackage{braket}
\usepackage{pifont}
\usepackage{blkarray, bigstrut}

\begin{document}
\title{Chimera states on m-directed hypergraphs}

\author{Rommel Tchinda Djeudjo}
\email{rommel.tchindadjeudjo@unamur.be}
\affiliation{Department of Mathematics \& naXys, Namur Institute for Complex Systems, University of Namur, B5000 Namur, Belgium}

\author{Timoteo Carletti}
\affiliation{Department of Mathematics \& naXys, Namur Institute for Complex Systems, University of Namur, B5000 Namur, Belgium}

\author{Hiroya Nakao}
\affiliation{Department of Systems and Control Engineering, Institute of Science Tokyo (former Tokyo Tech), Tokyo 152-8552, Japan}
\affiliation{International Research Frontiers Initiative, Institute of Science Tokyo (former Tokyo Tech), Kanagawa 226-8501, Japan}

\author{Riccardo Muolo}
\email{riccardo.muolo@riken.jp}
\affiliation{Department of Systems and Control Engineering, Institute of Science Tokyo (former Tokyo Tech), Tokyo 152-8552, Japan}
\affiliation{RIKEN Center for Interdisciplinary Theoretical and Mathematical Sciences (iTHEMS), Saitama 351-0198, Japan}

\date{\today}

\begin{abstract}

Chimera states are synchronization patterns in which coherent and incoherent regions coexist in systems of identical oscillators. This elusive phenomenon has attracted significant interest and has been widely analyzed, revealing several types of dynamical states. Most studies involve reciprocal pairwise couplings, where each oscillator exerts and receives the same interaction from neighboring ones, thus being modeled via symmetric networks. However, real-world systems often exhibit non-reciprocal, non-pairwise (many-body) interactions. Previous studies have shown that chimera states are more elusive in the presence of non-reciprocal pairwise interactions, while they are easier to observe when the interactions are reciprocal and higher-order (many-body).  In this work, we investigate the emergence of chimera states on non-reciprocal higher-order structures, called $m$-directed hypergraphs, {which we compare with their corresponding networks}, and we observe that {some types of chimera states} can emerge due to directionality, which had not been previously observed in {its absence}. We also compare the effect of non-reciprocal interactions between higher-order and pairwise couplings, and {we find numerically that chimera states appear over a broader parameter range when considering higher-order interactions than in the corresponding network case, demonstrating the impact of directionality and the effect of higher-order interactions}. Finally, the nature of phase chimeras has been further validated through phase reduction theory.

\end{abstract}

\maketitle

%--------------------------------
\section{Introduction}

Synchronization is a phenomenon observed in a wide variety of natural and engineered systems, ranging from flashing fireflies and cardiac pacemaker cells to power grids and neural circuits \cite{pikovsky2001synchronization}. It arises when individual self-sustained oscillatory units adjust their rhythms due to their interactions, i.e., the coupling, which is often modeled through a network \cite{arenas2008synchronization}. The structure of the interactions can lead to different kinds of synchronization patterns \cite{boccaletti2006complex}. Among the most peculiar patterns of synchronization, one can find a {\em chimera state}, a counterintuitive state in which coherent and incoherent behaviors coexist within the system of identical oscillators. The phenomenon was first reported by Kaneko in the context of coupled maps~\cite{kaneko1984period,kaneko}, and later observed in a variety of numerical studies involving both global, i.e., all-to-all~\cite{hakim1992dynamics,nakagawa1993collective,chabanol1997collective} and nonlocal~\cite{kuramoto1995scaling,kuramoto1996origin,kuramoto1997power,kuramoto1998multiaffine,kuramoto2000multi}, i.e., first neighbors, coupling schemes. Despite these earlier observations, the scientific community nowadays acknowledges the seminal work by Kuramoto and Battogtokh \cite{kuramoto2002coexistence} as the first systematic investigation and characterization of chimera states. This last work gained popularity after Abrams and Strogatz \cite{abrams2004chimera} coined the term ``chimera'' to describe the coexistence of different dynamical behaviors, inspired by the mythological creature composed of parts of different animals. Chimera states have also been identified in a range of experimental settings, lending credence to their physical relevance. These include Josephson junction arrays \cite{cerdeira_prl}, electronic circuits \cite{vale_chimeras,gambuzza2020experimental}, lasers \cite{hagerstrom2012experimental}, mechanical oscillators \cite{martens2013chimera}, and nano-electromechanical systems \cite{matheny2019exotic}. { Chimera states are transient (for finite systems) and highly elusive: in fact, except for some particular network topologies which make the chimera state robust~\cite{bram_malb_chim,muolo2024persistence,asllani2025pattern}, they are strongly dependent on initial conditions, parameters, and structure of the interactions, and even a small variation of any of those factors can cause their disappearing. For this reason,} considerable effort has been devoted to identifying configurations, e.g., specific ranges of parameters, coupling strengths, and network topologies, allowing the emergence and persistence of such patterns \cite{zakharova2020chimera,parastesh2021chimeras}. Scholars have shown particular interest in the framework of neuroscience, where chimera-like patterns have been suggested as models for unihemispheric sleep observed in certain animals \cite{chimera_neuro,majhi2019chimera,rattenborg2000behavioral}. Since most real-world networks, including brain networks, are strongly non-reciprocal \cite{malbor_teo}, the study of chimera states in the presence of non-reciprocal interactions becomes particularly relevant. However, except for a few works \cite{bick2015controlling,vasudevan2015earthquake,deschle2019directed,jaros2021chimera,bram_malb_chim,muolo2024persistence}, the vast majority of the study on chimera states, including the references mentioned above, assumes reciprocal (i.e., symmetric) coupling.

Nonetheless, considering non-reciprocal interactions is only a first step towards more realistic settings. In fact, in recent years, more complex structures, such as hypergraphs and simplicial complexes, have triggered the interest of scholars and allowed to move beyond the network framework \cite{battiston2020networks,battiston2021physics,bianconi2021higher,boccaletti2023structure,bick2023higher,muolo2024turing,millan2025topology}. This is because networks, despite being a very good approximation, do not always fully capture the interactions in complex systems, which are often not only pairwise, i.e., one-by-one, but rather higher-order, i.e., group, interactions \cite{battiston2020networks,battiston2021physics}. {Evidences come} from neuroscience \cite{rosen1989,wang2013,petri2014homological,sizemore2018cliques}, ecology \cite{grilli_allesina,iacopini2024not}, or social behaviors \cite{centola2018experimental}, to name a few. For what concerns {dynamical aspects}, higher-order interactions have been found to have great effects on the dynamics, for instance, in random walks \cite{carletti2020random,schaub2020random}, synchronization \cite{tanaka2011multistable,skardal2020higher,millan2020explosive,gambuzza2021stability,leon2024higher}, contagion \cite{iacopini2019simplicial,deville2020consensus}, or pattern formation \cite{carletti2020dynamical,muolo2023turing}. 

Chimera states have also been studied in presence of higher-order interactions, {which have been shown to enhance their emergence and persistence} \cite{kundu2022higher,ghosh_chimera2,bick_nonlocal1,muolo2024phase,muolo2025pinning}. In this work, we go one step forward by considering non-reciprocal, i.e., directed, higher-order interactions, whose effects on the dynamics have been studied in the context of synchronization of chaotic oscillators \cite{gallo2022synchronization,della2023emergence} and pattern formation \cite{dorchain2024impact}. In particular, we study the emergence of chimera states on $m$-directed hypergraphs \cite{gallo2022synchronization} and show the emergence of amplitude-mediated chimeras \cite{sethia2013amplitude} and phase chimeras \cite{zajdela2025phase}. Our results suggest that directionality {may be linked to the presence of chimera states, with indications that this effect becomes more noticeable in the presence of higher-order interactions. This can be observed when comparing the behavior on the directed clique-projected network, which is a possible representation of the corresponding pairwise structure.}
%which is weighted pairwise connection with all the nodes part of the same hyperedge
Furthermore, we validate the numerical results on phase chimeras by means of the phase reduction, an established technique in the study of oscillatory systems~\cite{nakao2016phase,kuramoto2019concept,pietras2019network}.

% support the claim that we are clearly facing a higher-order effect, enhanced by directed hypergraph and significantly fading away once directed clique-projected networks are considered

In the next Section, we introduce $m$-directed hypergraphs, the dynamical model and some characterization tools for the chimera states. The results of amplitude-mediated chimeras and phase chimeras, including the discussion on phase reduction, can be found, respectively, in Secs. \ref{sec:3} and \ref{sec:4}, right before the Conclusions.

\section{The framework}\label{sec:2}

In this Section, we first introduce the directed higher-order topology we will consider in our numerical study, together with its non-reciprocal pairwise counterpart; then, we present the model under study, the celebrated Stuart-Landau oscillator, and some quantities which are useful to characterize the kind of chimera state we are observing.

\subsection{m-directed hypergraphs and clique-projected networks}

A hypergraph is defined by the pair \((V, E) \), where \(V=\left\{ {v_1,v_2,...,v_N} \right\}\) is a set of nodes (or vertices) and \( E=\left\{ { {e_1 }  , {e_2 },..., {e_m }} \right\}\) is a set of hyperedges, each hyperedge being a subset of \(V\) containing at least two nodes. Note that a hyperedge composed of $2$ nodes is a pairwise link, and that a hypergraph with only such hyperedges is a network. A hypergraph is said to be \(k\)-uniform when each hyperedge consists of \(k\) nodes, which is the kind we will consider throughout this work. Moreover, the higher-order structures we consider are non-reciprocal, namely, $m$-directed hypergraphs \cite{gallo2022synchronization}. A hypergraph with a $d$-hyperedge, i.e., formed by \(d+1\) nodes~\footnote{Note that ,in some earlier works on hypergraphs, a $d$-hyperedge encodes a $d$-body interaction, while we follow the more common terminology, consistent with the literature on simplicial complexes, where a $d$-simplex encodes $(d+1)$-body interactions \cite{bianconi2021higher}.}, is said to be $m$-directed \((m\leq d)\) if the nodes can be split into two groups, one containing $m$ head nodes and the second one the remaining $q=d+1-m$ nodes, named tail nodes. The relevant fact is that each tail node influences each head node but the contrary is not true. On the other hand, head nodes interact among themselves. Note that this is a generalization of directed {networks}, where there is one head node affected by one tail node through a directed link. In this case, the adjacency tensor becomes the adjacency matrix \( A_{i,j} \), with the convention that it takes value $1$ (or a positive real number, if the network is weighted) when there is a directed link from node \( j \) to node \( i \), and 0 otherwise.

Let us consider a hyperedge consisting of the head group \(i_{1},\dots,i_{m}\) and the tail group \(j_{1},\dots ,j_{q}\); because head nodes interact with each other and nodes in the tail
interact interchangeably with nodes in the head, the $m$-directed $d$-hyperedge, $d=m+q-1$, has the following property
\begin{equation}
A_{\pi_1(i_1 \ldots i_m)\, \pi_2 (j_1  \ldots j_q )}^{(d)}  = 1\,,
\end{equation}
where $A^{(d)}$ is the $d$-th order adjacency tensor encoding $(d+1)$-body interactions, \(\pi_1(i_1 \ldots i_m)\) represents any permutation of the indices \(i_{1}, \ldots, i_{m}\), and \(\pi_2(j_1\ldots j_q)\) any permutation of the indices \(j_{1}, \ldots, j_{q}\). Note that the tensor can take a positive real value in the case of weighted hypergraph.

\begin{figure*}[ht!]
    \centering
     \includegraphics[scale=0.2]{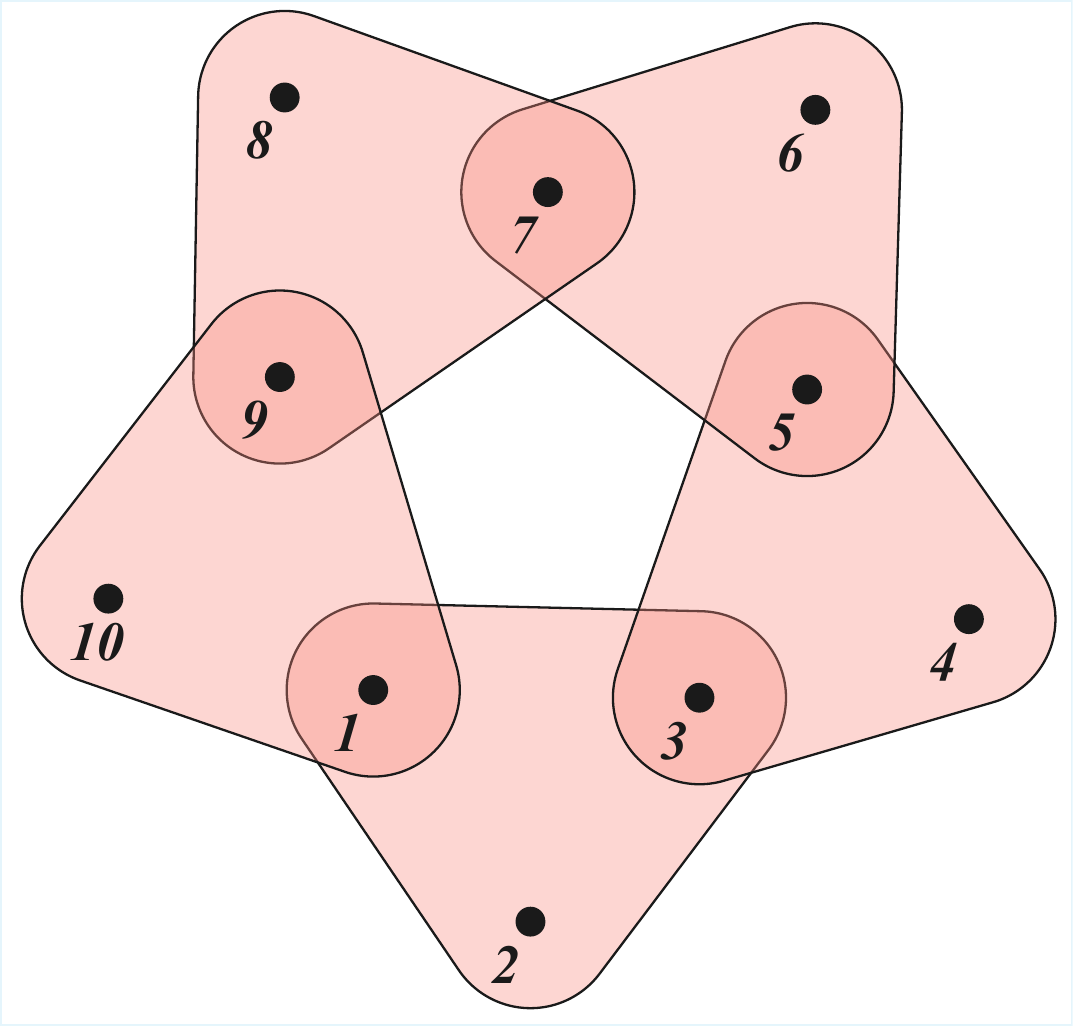}
    \caption{{An example of symmetric nonlocal $2$-hyperring with $10$ nodes, as introduced in \cite{muolo2024persistence}.}}
    \label{symetrie_hypergraph}
\end{figure*}

\begin{figure*}[ht!]
    \centering
    \includegraphics[width=0.9\textwidth]{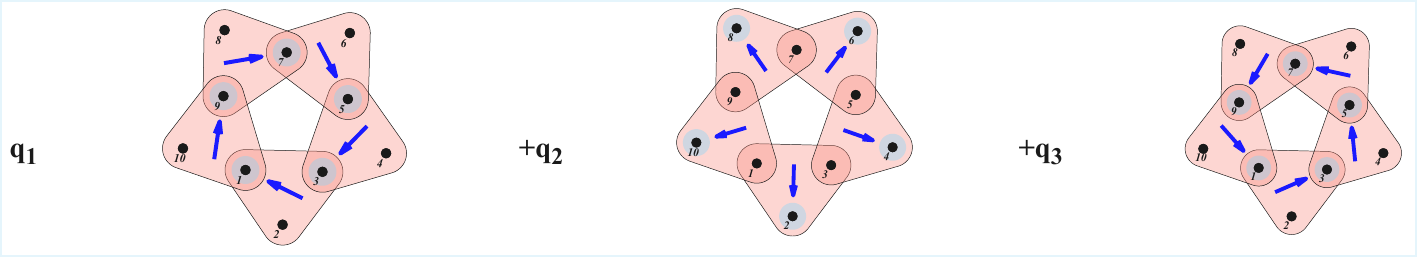}
\caption{We schematically represent a family of \textbf{$1$-directed $2$-hyperrings} obtained as a weighted ``combination'' of three base hyperrings. Each hyperring is made of $1$-directed $2$-hyperedges, weighted by the coefficients 
\( (q_1, q_2, q_3) \). When \( q_1 = q_2 = q_3 = 1\), the resulting structure is a \textbf{symmetric $2$-hyperring}, as shown in Fig.~\ref{symetrie_hypergraph}. On the other hand, different weights allow to generate a family of \textbf{$1$-directed $2$-hyperrings}, where some directions are favored. The heads of the hyperedges are highlighted in light blue, while arrows help to identify the directionality of the hyperedges, illustrating thus the direction of the interactions in the hyperring. In Appendix \ref{appB}, we show an analogous construction for the case of \textbf{$2$-directed $2$-hyperrings}.}

    \label{hypergraphe_1_diridé_2_hyperring}
\end{figure*}

In the present work, we consider the case of directed {nonlocal} hyperrings, which is a particular case of uniform hypergraph. {Symmetric nonlocal $d$-hyperrings (encoding $(d+1)$-body interactions) have been introduced in~\cite{muolo2024phase} and they model a ring-like higher-order structure where each hyperedge is adjacent to two hyperedges with whom it shares a node. An example of such structure, namely, a $2$-hyperring,} is reported in Fig.~\ref{symetrie_hypergraph}, whereas in Fig.~\ref{hypergraphe_1_diridé_2_hyperring} we can appreciate a $1$-directed $2$-hyperring built by ``juxtaposing'' three weighted $1$-directed $2$-hyperedges. Let us observe that the chosen structure allows us to have a rotation invariance, often used in models of chimera states on networks, and it is a possible generalization of a higher-order nonlocal coupling \cite{muolo2024phase}, although not the only one (see, for example, Refs. \cite{bick_nonlocal1,bick_nonlocal2,zhang2024deeper}). In the following we will denote by ``junction'' nodes those elements shared among two adjacent hyperedges.

The proposed construction allows to obtain a family of hyperrings whose directionality can be controlled by tuning three parameters, $q_1$, $q_2$ and $q_3$. There are two methods of tuning the directionality \cite{gallo2022synchronization}. The first one, named method $(i)$, relies on fixing one weight, say $q_1=1$, and then changing the values of the remaining ones. For a $2$-hyperring, one can thus have
\[
q_1 = 1, \quad q_2 = q_3 = p, \quad \text{with } p \in [0,1]\, .
\]
This method does not conserve the total weight of the hyperedge and, thus, the (generalized weighted) degree is not preserved neither. An alternative method, called method $(ii)$, consists in normalizing the parameters $q_1+q_2+q_3=1$, preserving thus the total weight of the hyperedge. For a $2$-hyperring, we have
\[
q_1 = 1 - 2p, \quad q_2 = q_3 = p, \quad \text{with } p \in \left[0,\frac{1}{3}\right]\, ,
\]
in such a way that a symmetric hyperring can be recovered for $p=1/3$. The results discussed in Section \ref{sec:results} are for method $(i)$, but there is no qualitative difference when we use method $(ii)$ to tune the directionality {(data not shown).}

\begin{figure*}[ht!]
    \centering
    \includegraphics[width=0.7\textwidth]{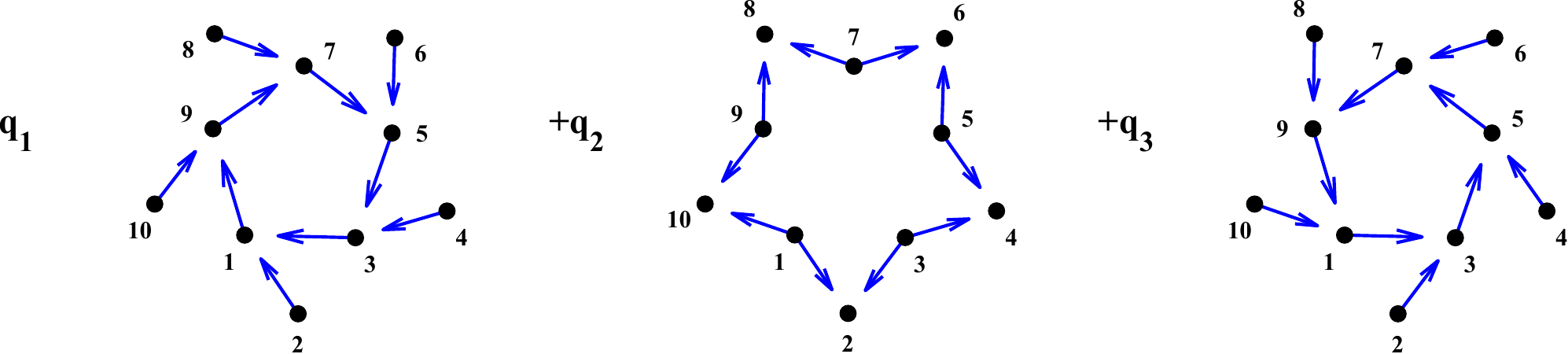}
    \caption{We schematically represent a family of clique-projected networks, obtained from the $1$-directed $2$-hyperring presented in Fig.~\ref{hypergraphe_1_diridé_2_hyperring}. Each directed clique is obtained from the directed hyperedge and it is weighted by the coefficients \( (q_1, q_2, q_3) \). When \( q_1 = q_2 = q_3 = 1\), the resulting structure is a symmetric network.
}
    \label{fig:clique_projection_1}
\end{figure*}

Lastly, we introduce a network obtained from the directed hypergraph, {which we use as a reference to qualitatively assess the emergence of chimera states in comparison with the higher-order case. Specifically,} a directed clique-projected network
obtained by adding a directed link between nodes in the tail of the hyperedge and nodes in the head of the hyperedges. Nodes in the head will also be connected among themselves via symmetric links.  In general, the resulting directed clique-projected network will be weighted. An example of such a structure is shown in Fig.~\ref{fig:clique_projection_1}. {As explained in~\cite{muolo2024phase}, let us stress that the higher-order system and its clique-projected counterpart are intrinsically different and not directly comparable in a strict sense. In particular, the higher-order and pairwise formulations are intrinsically different, and the projection procedure results in an interaction structure that cannot fully capture the underlying higher-order dynamics. Lastly, let us observe that the clique-projected network adopted here is only one possible choice of the projection. Other representations are possible, but have not been found to affect the emergence of chimera states~\cite{muolo2024phase}.}

\subsection{Stuart-Landau oscillators}

Let us now consider \(N\) identical nonlinear systems coupled via a $1$-directed hypergraph of order \(D\), with \(D \geqslant 2\). Because of the presence of a single node in the head, we will denote the $d$-hyperedge by $A_{ij_1  \ldots j_q }^{(d)}$. The dynamics is given by the following set of equations
\begin{equation}
\dot{\vec{x}}_{i} = \vec{f}(\vec{x}_i) + \sum_{d = 1}^D c_d \sum_{j_1, \dots, j_d = 1 }^N 
A_{i j_1 j_2 \dots j_d }^{(d)} \vec{g}^{(d)} (\vec{x}_i ,\vec{x}_{j_1} ,\dots,\vec{x}_{j_d})\, ,
\end{equation}
where, for all $i=1,\dots,N$, \(\vec{x}_i \in \mathbb{R}^m\) denotes the state vector describing the dynamics of the $i$-th oscillator, \(\vec{f}: \mathbb{R}^m \to \mathbb{R}^m\) is the nonlinear function determining the evolution of the system, and \(\vec{g}^{(d)}: \mathbb{R}^{m \times (d+1)} \to \mathbb{R}^m\) represents the nonlinear function responsible for the coupling among nodes in the same hyperedge. Note that, for us $m=2$, as the system introduced below has such dimensionality. We hereby assume the coupling to be diffusive--like, namely
\begin{eqnarray}
\vec{g}^{(d)} (\vec{x}_i ,\vec{x}_{j_1} ,...,\vec{x}_{j_d}) &=& \vec{h}^{(d)} (\vec{x}_{j_1} ,\vec{x}_{j_2} ,...,\vec{x}_{j_d}) \nonumber \\
&-& \vec{h}^{(d)} (\vec{x}_i ,\vec{x}_i ,...,\vec{x}_i),
\end{eqnarray}
where \(\vec{h}^{(d)} :\mathbb{R}^{m \times d}  \to \mathbb{R}^m\). Then, the previous equations of motion become
\begin{align}
\dot{\vec{x}}_i &= \vec{f}(\vec{x}_i) + \sum_{d=1}^D c_d \sum_{j_1, j_2, \dots, j_d = 1}^N A_{i j_1 j_2 \dots j_d}^{(d)} \nonumber\\
&\quad \times \left[ \vec{h}^{(d)}(\vec{x}_{j_1}, \vec{x}_{j_2}, \dots, \vec{x}_{j_d}) 
- \vec{h}^{(d)}(\vec{x}_i, \vec{x}_i, \dots, \vec{x}_i) \right],
\end{align}
with $A_{i \pi_2 ( j_1 \ldots j_d )}^{(d)}$ keeping the same value for any permutation $\pi_2$, the hypergraph being $1$-directed. In the case of $1$-directed $2$-hyperrings, the coupling functions are chosen to be
\begin{equation}
\vec{h}^{(2)}(\vec{x}_{j_1},\vec{x}_{j_2})=\begin{bmatrix}x_{j_1}^2 x_{j_2} \\ x_{j_1}^2 x_{j_2} \end{bmatrix}\quad \text{ and }\quad \vec{h}^{(2)}(\vec{x}_i, \vec{x}_i) = \begin{bmatrix} x_i^3 \\ x_i^3 \end{bmatrix}\, .
\end{equation}

We consider a system made of $N$ interacting Stuart-Landau units, a paradigmatic model in the study of synchronization dynamics, as it corresponds to the normal form of the supercritical Hopf-Andronov bifurcation~\cite{nakao2014complex}. With the coupling as above, the equations read

\begin{equation}\label{eq:ho_SL}
\left\{
\begin{aligned}
   \frac{dx_i}{dt} &=& \alpha x_i - \omega y_i - \left( x_i^2 + y_i^2 \right) x_i   +\epsilon \sum_{j_1,\dots,j_2}  A^{(1)}_{i, j_1,j_2}( x_{j_1}^2 x_{j_2} - x_i^3) , \\
   \frac{dy_i}{dt} &=& \omega x_i + \alpha y_i - \left( x_i^2 + y_i^2 \right) y_i +\epsilon \sum_{j_1,\dots,j_2}  A^{(1)}_{i, j_1,j_2}( x_{j_1}^2 x_{j_2} - x_i^3 ),
\end{aligned}
\right.
\end{equation}

\noindent where \( \alpha \) is a bifurcation parameter and \( \omega \) is the frequency of the oscillators. Note that the coupling follows the configuration $x\rightarrow x,y\rightarrow x$, as in previous works \cite{gambuzza2020experimental,muolo2024phase,muolo2025pinning}. Let us observe that units are identical, namely, the parameters \( \alpha \) and \( \omega \) are the same for every unit in the system. Each isolated system exhibits a stable limit cycle for \( \alpha > 0 \), condition that we assume true throughout this study. 

 The dynamics of corresponding system on the clique-projected network is given by
\begin{equation}\label{eq:clique_SL}
\left\{
\begin{aligned}
    \dot{x}_i &= \alpha x_i - \omega y_i - \left( x_i^2 + y_i^2 \right) x_i + \epsilon \sum_{j = 1}^N A_{i, j} \left( x_{j}^3 - x_i^3 \right), \\
    \dot{y}_i &= \omega x_i + \alpha y_i - \left( x_i^2 + y_i^2 \right) y_i + \epsilon \sum_{j = 1}^N A_{i, j} \left( x_{j}^3 - x_i^3 \right)\,. 
\end{aligned}
\right.
\end{equation}

\subsection{Frequency, phase, and amplitude}\label{subsec:charact_tool}

Before presenting the results, let us introduce the { metrics} we will use to characterize chimera states. After a transient interval and for a given time window, the orbit of the $i$-th oscillator can be written as ${ a_i}\exp[\imath(2\pi\Omega_i t + \theta_i)]$.  {By performing a Fourier analysis over the considered time window, we identify the dominant peak in the power spectrum, which defines the frequency $\Omega_i$ of the oscillator, together with the associated amplitude $a_i$ and phase $\theta_i$.} To be more precise {in the definition of the terms}, $\omega=2\pi\Omega$ is the \textit{angular frequency}, whilst $\Omega$ is the properly called \textit{frequency}; in what follows we will call them both "frequency", as we believe that there is no ambiguity and it is clear what they mean. { Let us observe, that the reconstructed phase, amplitude and frequency will not depend on the time window only if the signal is periodic. On the contrary, to each time window we can associate a value of $a_i$, $\Omega_i$ and $\theta_i$, $i=1,\dots,N$, and thus compute the average of the involved quantity over the used time windows, denoted by $\langle \cdot \rangle$. More precisely, we performed the Fourier analysis on the resulting signals restricted to the time interval $[900,1000]$. Then, the latter has been further subdivided into $n$ sub-intervals, in each of which we extracted information on $\omega^{(j)}_i$, $a^{(j)}_i$, and $\theta^{(j)}_i$ for each oscillator $i$ and $j=1,\dots,n$, to eventually compute the average and the standard deviation :
\begin{equation}
    \label{eq:ave}
    \langle Z_i\rangle = \frac{1}{n}\sum_{j=1}^n Z_i^{(j)}\text{ and } \sigma^2(Z_i) = \frac{1}{n}\sum_{j=1}^n \left(Z_i^{(j)}-\langle Z_i\rangle\right)^2\, ,
\end{equation}
where $Z\in\{a,\Omega,\theta\}$. In the following figures, the averages are displayed as blue dots and the standard variations as shaded light blue regions. Let us observe that the standard deviation is a measure of temporal coherence of the signal on a given node and thus, in the present framework, it is related to the wave-like behavior of the chimera state.}

Indeed, when the chimera behavior, i.e., the coexistence of {spatial} coherence and incoherence, is relative to amplitude and this induces a chimera behavior also on the phase, we talk about \textbf{amplitude-mediated chimeras} \cite{sethia2013amplitude}. When the chimera behavior is only with respect to the phase, while amplitudes and frequencies are constant, then we are dealing with \textbf{phase chimeras} \cite{zajdela2025phase}. 

{ In order to summarize the previous considerations and to give a proper characterization, we make use of the notion of {\em normalized total variation}, a concept originating from analysis and previously applied to the study of chimera states \cite{muolo2024phase}. For each of the quantities introduced above, its variation is defined as follows: 
{\begin{equation}
\label{eq:totnormV}
\begin{cases}
\displaystyle
    V(\langle a \rangle) = \frac{1}{N} \sum_{i=1}^N |\langle a_{i+1} \rangle - \langle a_i\rangle|\, \\
    \displaystyle
V({\langle \omega \rangle }) = \frac{{2\pi}}{N} \sum_{i=1}^N |\langle \Omega_{i+1} \rangle - \langle \Omega_i \rangle|, \\ 
\displaystyle V(\langle \theta \rangle) = \frac{1}{\pi N} \sum_{i=1}^N \lVert \langle \theta_{i+1} \rangle - \langle \theta_i \rangle \rVert\, , 
\end{cases}
\end{equation}}
where the circular distance is given by $\lVert \theta \rVert = \min\{\theta, 2\pi - \theta\}$ for any $\theta \in [0,2\pi)$. Indices are taken modulo $N$, meaning $N+1 \equiv 1$. 

The normalized total {variation} evaluates how smoothly a function behaves: a small variation indicates regularity, whereas a large value ($<1$, due to the normalization) highlights abrupt transitions between neighboring points. Consequently, if the normalized total phase variation, $V(\langle \theta \rangle)$, is large while the amplitude and { angular frequency} normalized total variations are negligible, $V(\langle a \rangle)\sim 0$ and $V({\langle \omega \rangle})\sim 0$, the system displays a phase chimera. On the other hand, when $V(\langle \theta \rangle)\sim 0$ and $V({\langle \omega \rangle})\sim 0$ but the normalized total amplitude variation, $V(\langle a\rangle)$, is large, the outcome corresponds to an amplitude {chimera\footnote{Let us observe that in the first paper where amplitude chimeras were introduced~\cite{zakharova2014chimera,zakharova2020chimera}, authors did not distinguish between phase and angular frequency terms and thus defined such new state only according to the amplitude.}.} {All these quantities, although being a useful tool in the characterization of chimera states, do not provide a clear distinction between what is a chimera state and what is not, but rather provide a quantitative measure of the ``smoothness" in the variation.} Finally, it is important to clarify that the normalized total variation cannot easily distinguish between amplitude-mediated chimeras and incoherent states{, as it aggregates local information into a single scalar quantity. While it provides a useful quantitative measure to support the analysis, it does not by itself offer a definitive characterization of the observed patterns. Therefore, complementary approaches, such as direct visualization of the system dynamics, are necessary to properly interpret the different regimes.}

% {when for instance all the normalized total variation are too much large we have an incoherent state.}

% {It should also be noted that for variations close to 1, we have an inconsistent state.}

%{We also note that, in the context of this work,  we performed Fourier analysis on the resulting signals restricted to the window $[900,1000]$. This window was subdivided into $n$ sub-intervals, in each of which we extracted information on $\omega$, $a$, and $\theta$ for each oscillators. Finally, we computed the average of the parameters obtained over each sub-interval. Note that the symbol $\langle \cdot \rangle$ denotes the average.}

\section{Results}
\label{sec:results}
The aim of this section is to present our main results obtained by numerically integrating Eqs.~\eqref{eq:ho_SL} starting from clustered initial conditions, \(x_{i} = 1, y_{i} = -1\) for \(i \in [1, N/2]\), and \(x_{i} = -1, y_{i} = 1\) for \(i \in [N/2 + 1, N]\){, i.e., half of the oscillators are in anti-phase configuration with respect to the other half. {Let us note that this setting for the initial conditions has been used in several studies about chimera states \cite{zakharova2014chimera,muolo2024phase,muolo2025pinning}.} We call such state \textit{coherent clusters}, to distinguish it from the \textit{coherent state} in which  phases vary smoothly between adjacent nodes. Note that the directed hyperrings, and, as a consequence, also the clique-projected networks, possess some symmetry properties. If we number the nodes in a regular way, e.g., counter-clock wise, it does not matter which nodes are in the first or the second cluster, as long as they are consecutive.} In what follows, we will vary the directionality of the hyperedges by using the method not preserving the coupling strength of the hyperedge (i.e., method $(i)$ of the previous Section). We will fix a privileged direction by setting $q_1=1$ and then uniformly vary $q_2 = q_3 = p$. When $p=1$, the hypergraph is symmetric, while the asymmetry is introduced when $1>p\geq0$. The results obtained with method $(ii)$ show no significant difference, hence, we will not show them. {In both cases, the observed chimera states {seem to be} enhanced by the presence of directed higher-order interactions, {as denoted by higher values of the variations. This further corroborates} previous results~\cite{kundu2022higher,muolo2024phase,muolo2025pinning}.}

{ All simulations have been performed with the Runge-Kutta IV order explicit integration method with integration step $0.01$ and by using the software Matlab \cite{Matlab}.} In the Appendices, we show additional results for a different orientation of $1$-directed $2$-hyperring (Appendix \ref{appA}) and for $2$-directed $2$-hyperrings (Appendix \ref{appB}).

\subsection{Amplitude-mediated chimera states}\label{sec:3}

\begin{figure*} [ht!]
			\includegraphics[width=1\textwidth]{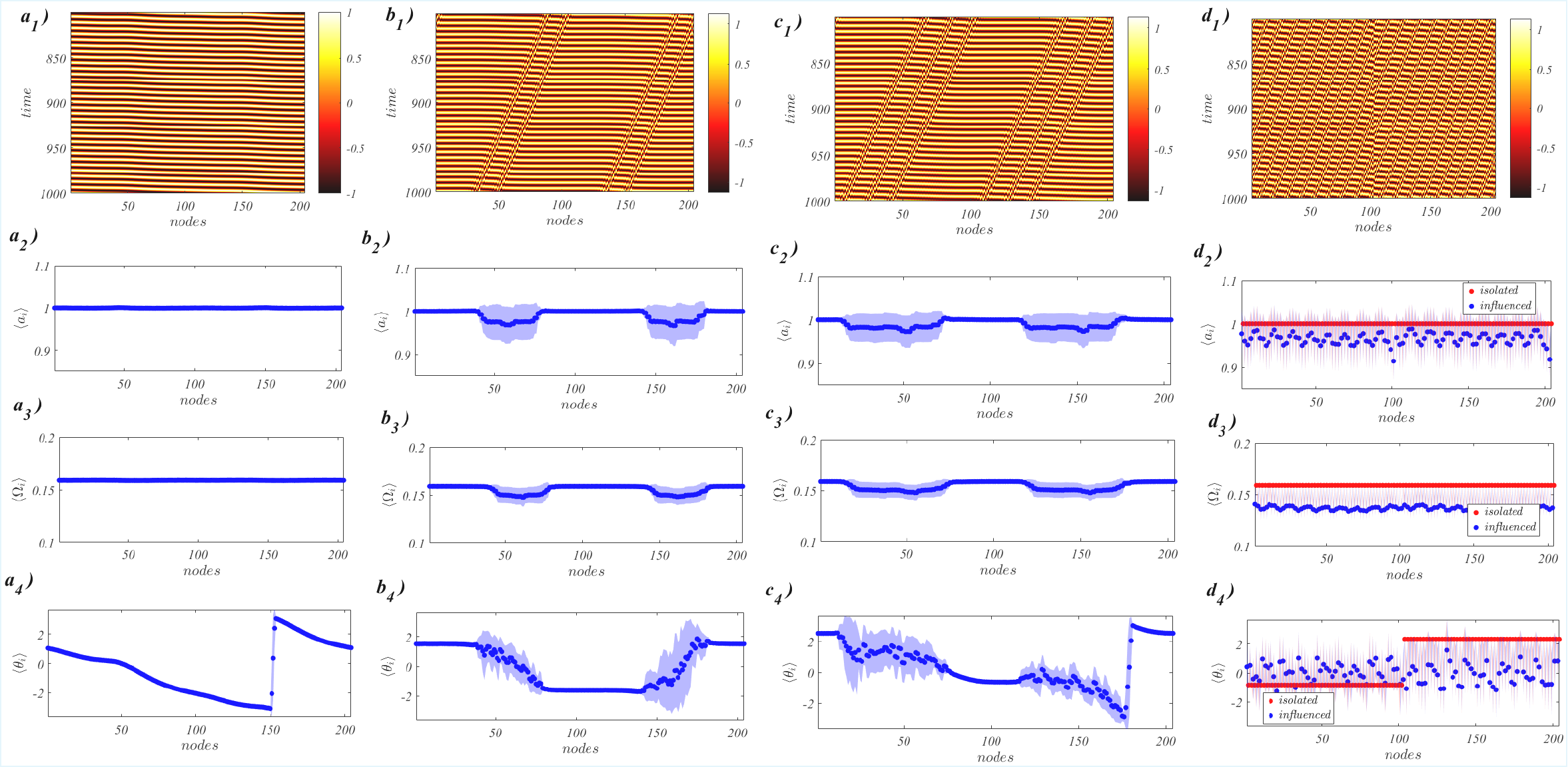} 
		 \caption{Analysis of the dynamics on a $1$-directed $2$-hyperring of $204$ nodes. The first row shows the spatiotemporal diagrams {for the $y$ variable (the behavior of the $x$ variable is analogous)}, the second row the {average} amplitudes, the third row the {average} frequencies, and the last row the {average} phases. The directionality parameter $p$ is varied with the columns: 
(a1, a2, a3, a4) coherent behavior for \( p = 1 \) {(with $V(\langle a \rangle) \approx 2.45\, {\times}10^{-4}$, $V(\langle \omega \rangle) = 1.28 \, {\times}10^{-3}$, and $V(\langle \theta \rangle) \approx 1.92 \, {\times}10^{-2}$)} ;  
(b1, b2, b3, b4)  traveling amplitude-mediated chimera state for \( p = 0.2 \) { (with  $V(\langle a \rangle) \approx 0.0125$, $V(\langle \omega \rangle) = 0.025$, and $V(\langle \theta \rangle) \approx 0.076$)};  
(c1, c2, c3, c4)  traveling amplitude-mediated chimera state for \( p = 0.1 \) {(with $V(\langle a \rangle) \approx 0.0148$, $V(\langle \omega \rangle) = 0.044$, and $V(\langle \theta \rangle) \approx 0.201$)};  
(d1, d2, d3, d4) incoherent behavior for \( p = 0 \) {(with $V(\langle a \rangle) \approx 0.0348$, $V(\langle \omega \rangle) = 0.138 $, and $V(\langle \theta \rangle) \approx 0.527$)}. The model parameters are $\alpha=1$ and $\omega=1$, and the coupling strength is \( \epsilon = 0.2 \). {The shaded light blue area represents the standard deviation of the quantity under scrutiny, computed over $10$ consecutive sub-intervals, and quantifies the temporal variability of the node dynamics around its mean value.} }
    \label{f11}
	\end{figure*}

\begin{figure*} [ht!]
			\includegraphics[width=1\textwidth]{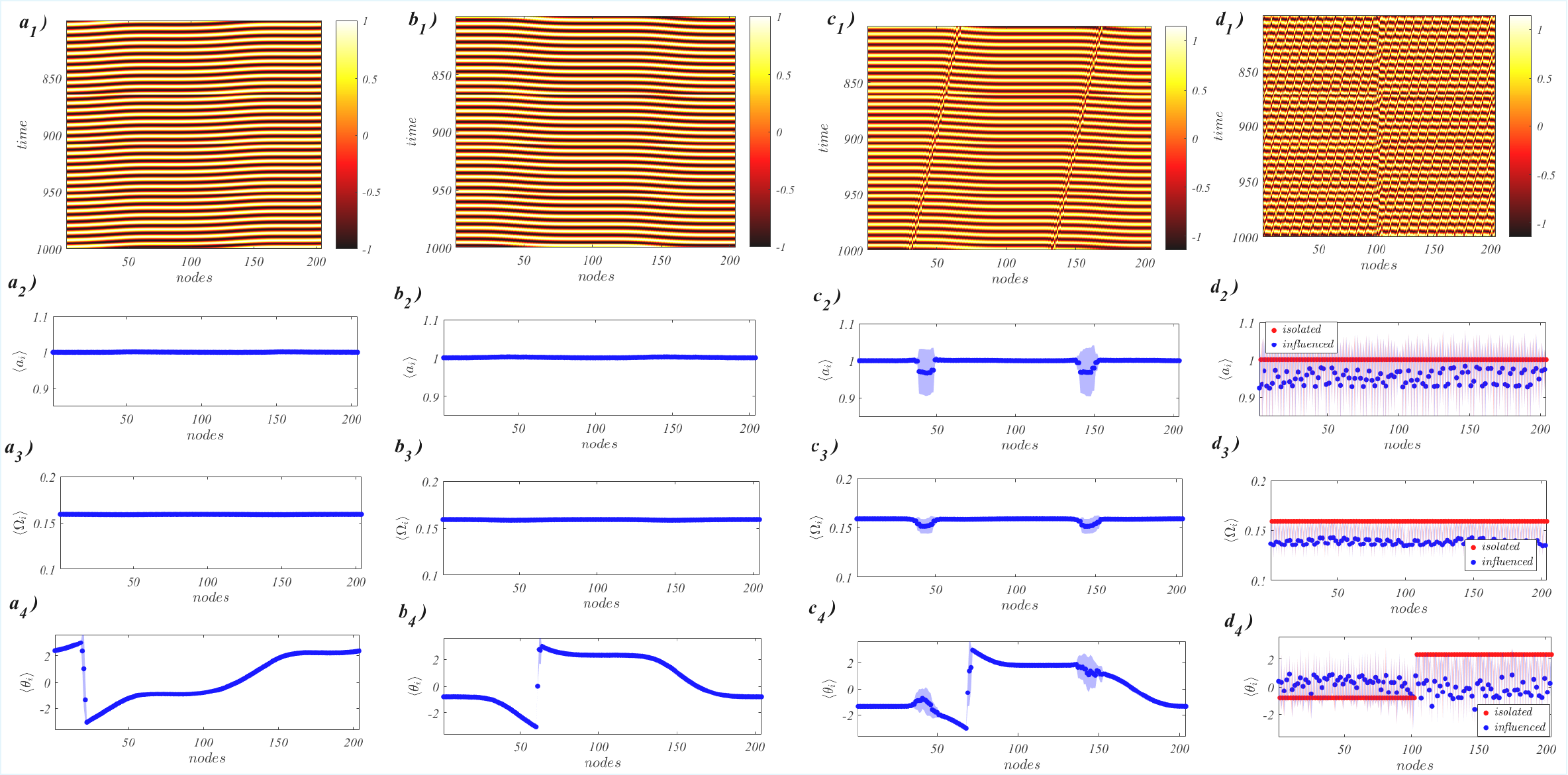} 
\caption{Analysis of the dynamics on a clique-projected network of $204$ nodes. The first row shows the spatiotemporal diagrams {for the $y$ variable (the behavior of the $x$ variable is analogous)}, the second row the {average} amplitudes, the third row the {average} frequencies, and the last row the {average} phases. The directionality parameter $p$ is varied with the columns: 
(a1, a2, a3, a4) coherent behavior for \( p = 1\) {(with $V(\langle a \rangle) \approx 5.0\, {\times}10^{-4}$, $V(\langle \omega \rangle) = 1.13\, {\times}10^{-4}$, and $V(\langle \theta \rangle) \approx 0.0185$)} ;  
(b1, b2, b3, b4)  coherent behavior for \( p = 0.2 \) {(with $V(\langle a \rangle) \approx 5.09\,{\times}10^{-4} $, $V(\langle \omega \rangle) = 2.32\,{\times}10^{-4}$, and $V(\langle \theta \rangle) \approx 0.0174$)};  
(c1, c2, c3, c4) traveling amplitude-mediated chimera  state for \( p = 0.1 \) 
{(with $V(\langle a \rangle) \approx 6.16\,{\times}10^{-3} $, $V(\langle \omega \rangle) = 0.008 $, and $V(\langle \theta \rangle) \approx 0.0571$)};  
(d1, d2, d3, d4)  incoherent behavior for \( p = 0 \) {(with $V(\langle a \rangle) \approx 0.0477$, $V(\langle \omega \rangle) = 0.1306 $, and $V(\langle \theta \rangle) \approx 0.5765$)}. 
The model parameters are $\alpha=1$ and $\omega=1$, and the coupling strength is \( \epsilon = 0.2 \). {The shaded light blue area represents the standard deviation of the quantity under study, computed over $10$ consecutive sub-intervals, and quantifies the temporal variability of the node dynamics around its mean value.} }

    \label{clique_projection_epsi_0.2}
	\end{figure*}

Let us first consider a configuration in which no chimera is observed when the topology is symmetric. In Fig.~\ref{f11} we show the effect of the directionality, tuned by the parameter \(p\). The reported results demonstrate that intermediate values of \(p\) can promote the emergence of chimera states, namely, \textbf{amplitude-mediated chimeras}. Moreover, such patterns are not stationary { amplitude-mediated chimeras, but traveling amplitude-mediated chimeras. For this reason, amplitudes, frequencies, and phases for each oscillator need to be computed with their standard deviation (in shaded blue in the Figures). In fact, over the time window in which we apply the Fourier method, the incoherence fronts are moving, meaning that a given node may belong to the incoherent region, or coherent one, only for a fraction of the time.} Starting with $p=1$, i.e., symmetric case (leftmost column), we observe a coherent behavior in the whole system (top row), where all oscillators have the same constant amplitude $a_i$ (second row from the top), same constant frequency $\Omega_i$ (third row from the top) and coherent phases, $\theta_i$ (bottom row). By decreasing $p$ (namely, from $p\lesssim 0.45$), i.e., by increasing the directionality, the system tends to favor the formation of complex dynamical patterns (see the second and the third columns from the left in Fig.~\ref{f11}). In particular, let us emphasize the amplitudes distributed in two groups, one for which $a_i$ are very close to $1$ and the remaining ones whose amplitudes are distributed in $({0.9},1)$. {From this observation we can observe that such traveling amplitude-mediated chimera states seem to emerge because of the directionality}. Let us, however, observe that when the directionality is maximal, namely $p=0$ (rightmost column), the chimera pattern vanishes, as it can be appreciated by looking at the rightmost column of Fig.~\ref{f11}. This fact can be explained by observing that, once $p=0$, because of the used regular structure, two types of nodes emerge: isolated nodes (non-junction nodes in this case) and non-isolated nodes (junction nodes). Each node in the former group oscillates on its own without receiving any input from any other node, meaning they act as ``external pacemakers'' towards nodes in the second group. The nodes of the first group, being isolated oscillators, follow their limit cycle solution with constant amplitude (see {red} dots in panel d2), constant frequency (see {red} dots in panel d3) and phases $\theta_i$ reminiscent of the initial conditions (see {red} dots in panel d4). Non-isolated nodes are constantly driven by the isolated ones and are not able to settle on any regular behavior; the amplitudes are close but irregularly distributed among the nodes (see {blue} dots in panel d2), the frequencies are constant but slightly different from the one of the limit cycle (see {blue} dots in panel d3), and, finally, the phases are randomly scattered in $(-\pi,\pi)$ (see {blue} dots in panel d4).

To test the role of higher-order interactions in the emergence of this behavior, we compared the previous results with the case of directed clique-projected networks. We thus numerically solved system~\eqref{eq:clique_SL} by using the same initial conditions, coupling strength, parameters and directionality tuning as above. The results are reported in Fig.~\ref{clique_projection_epsi_0.2} and show a different behavior. In the symmetric case, i.e., $p=1$, (leftmost column) and with $p=0.2$ (second column from the lest), the system exhibits a coherent behavior. For $p=0.1$ (second column from the right), amplitude-mediated chimeras appear; nonetheless, the incoherent region is small if compared to the same obtained in presence of many-body interactions. In particular, no traveling patterns are observed whatsoever. {As explained in the previous Section, an exact comparison of the pairwise and higher-order settings is not possible, given their intrinsically different nature. Nonetheless, we should remark the different qualitative behavior of the two setting, more specifically, that traveling patterns emerge only in the higher-order setting.} Finally, for $p=0$ (rightmost column), we have isolated (pacemakers--like) and non-isolated nodes, and the behavior is analogous to what is observed in the higher-order case.

The chimera patterns observed in this setting are interesting because they are caused by the combination of higher-order interactions and directionality. For high-directionality (i.e., low values of $p$) some kind of pattern would be expected, as the non-isolated nodes are forced oscillators, a setting that is known to give rise to chimeras \cite{muolo2024persistence}. However, thanks to the presence of higher order interactions, chimera states are observed also for intermediate values of directionality.

\subsection{Phase chimera states}\label{sec:4}

\begin{figure*} [ht!]
			\includegraphics[width=1\textwidth]{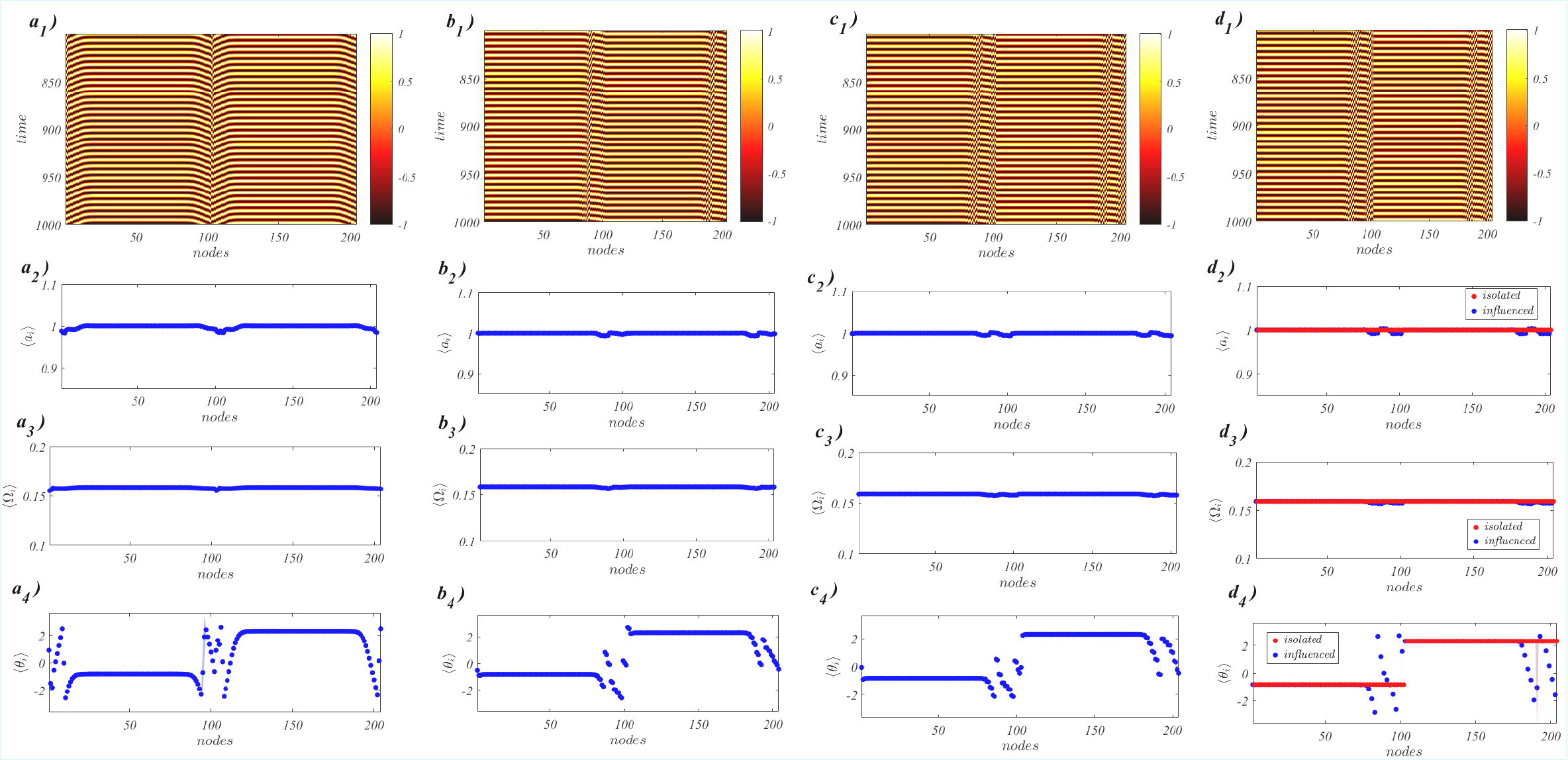} 
 \caption{
 Analysis of the dynamics on a $1$-directed $2$-hyperring of $204$ nodes. The first row shows the spatiotemporal diagrams {for the $y$ variable (the behavior of the $x$ variable is analogous)}, the second row the {average} amplitudes, the third row the {average} frequencies, and the last row the {average} phases. The directionality parameter $p$ is varied with the columns: 
(a1, a2, a3, a4) phase chimera state for \( p = 1 \) {(with $V(\langle a \rangle) \approx 1.91\,{\times}10^{-3} $, $V(\langle \omega \rangle) = 8.0\,{\times}10^{-4}$, and $V(\langle \theta \rangle) \approx 0.0659$)} ;  
(b1, b2, b3, b4)  phase chimera state for \( p = 0.2 \) {(with $V(\langle a \rangle) \approx 1.27\,{\times}10^{-3}$, $V(\langle \omega \rangle) = 1.49\,{\times}10^{-3}$, and $V(\langle \theta \rangle) \approx 6.57\,{\times}10^{-2}$)};  
(c1, c2, c3, c4) phase chimera state for \(p = 0.1\) {(with $V(\langle a \rangle) \approx 1.58\,{\times}10^{-3}$, $V(\langle \omega \rangle) = 0.0022$, and $V(\langle \theta \rangle) \approx 0.0995$)} ;  
(d1, d2, d3, d4) phase chimera state for \(p = 0\) {(with $V(\langle a \rangle) \approx 1.44\,{\times}10^{-3}$, $V(\langle \omega \rangle) = 2.6\,{\times}10^{-3}$, and $V(\langle \theta \rangle) \approx 0.1230$)}. The model parameters are $\alpha=1$ and $\omega=1$, and the coupling strength is \( \epsilon = 0.015\). {The shaded light blue area represents the standard deviation of the quantity under scrutiny, computed over $10$ consecutive sub-interval, and quantifies the temporal variability of the node dynamics around its mean value.} }
   \label{f1121}
	\end{figure*}

\begin{figure*} [ht!]
			\includegraphics[width=1\textwidth]{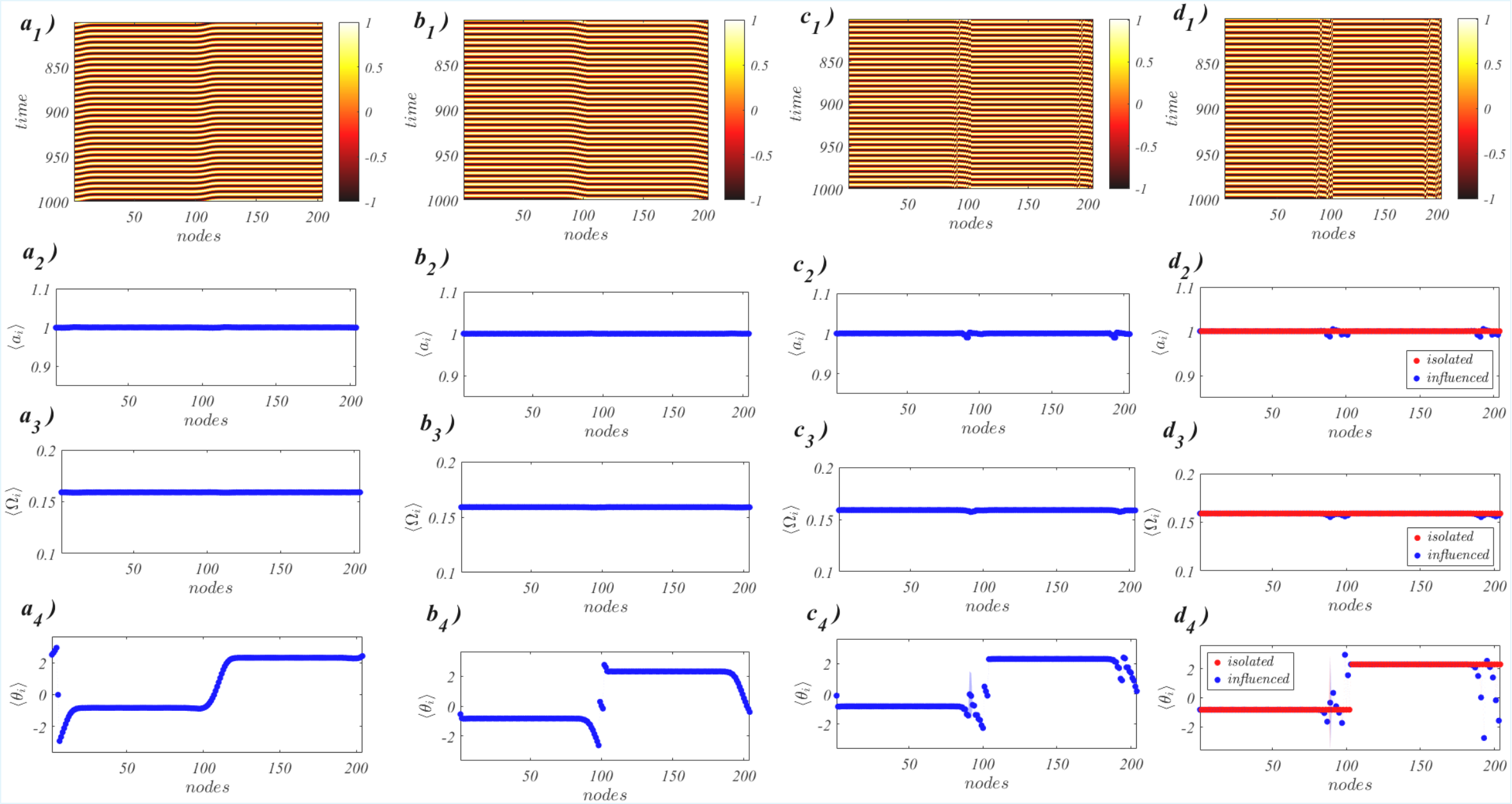} 
\caption{Analysis of the dynamics on a clique-projected network of $204$ nodes. The first row shows the spatiotemporal diagrams {for the $y$ variable (the behavior of the $x$ variable is analogous)}, the second row the {average} amplitudes, the third row the {average} frequencies, and the last row the {average} phases. The directionality parameter $p$ is varied with the columns: 
(a1, a2, a3, a4) coherent clusters for \( p = 1 \) {(with $V(\langle a \rangle) \approx 2.24\,{\times}10^{-4} $, $V(\langle \omega \rangle) = 6.28 \, {\times}10^{-5}$, and $V(\langle \theta \rangle) \approx 0.0103$)};  
(b1, b2, b3, b4)  coherent clusters for \( p = 0.2 \) {(with $V(\langle a \rangle) \approx 2.02\,{\times}10^{-4} $, $V(\langle \omega \rangle) = 1.0\,{\times}10^{-4}$, and $V(\langle \theta \rangle) \approx 0.0199$)};  
(c1, c2, c3, c4)  phase chimera state for \( p = 0.1 \) {(with $V(\langle a \rangle) \approx 8.93\,{\times}10^{-4} $, $V(\langle \omega \rangle) = 0.001$, and $V(\langle \theta \rangle) \approx 0.0479$ )};  
(d1, d2, d3, d4) phase chimera state for \( p = 0 \) {(with $V(\langle a \rangle) \approx 1.02\,{\times}10^{-3} $, $V(\langle \omega \rangle) = 0.002$, and $V(\langle \theta \rangle) \approx 0.0715$)}. 
The model parameters are $\alpha=1$ and $\omega=1$, and the coupling strength is \( \epsilon = 0.015 \). {The shaded light blue area represents the standard deviation of the quantity under consideration, computed over $10$ adjacent sub-intervals, and quantifies the temporal variability of the node dynamics around its mean value.} }
 
   \label{clique1}
	\end{figure*}

We now consider the same setting as before, but for a much smaller value of the coupling strength. In the higher-order setting, the symmetric case already  exhibits phase chimeras, i.e., the oscillators have (almost) identical amplitudes, $a_i\sim 1$ and identical frequencies $\Omega_i$, while we can observe the presence of coherent and incoherent domains for the phases $\theta_i$. The results are reported in Fig.~\ref{f1121}. It is interesting to observe that the chimera behavior persists when the directionality is introduced. 

When we compare those results with the pairwise case, i.e., clique-projected network, we remark that only for high directionality (i.e., small values of $p$) we appreciate some kind of phase chimera behavior, but the region of incoherence is smaller (see Fig.~\ref{clique1}).

As for the case shown in the previous section, some kind of patterns for high directionality would be expected. What is interesting is that the phase chimera behavior is conserved from the symmetric to the fully asymmetric case, {a result rooted on} the presence of higher-order interactions.

\subsubsection{Validation {of the phase chimera behavior} through phase reduction theory}

   \begin{figure*} [ht!]
			\includegraphics[width=0.98\textwidth]{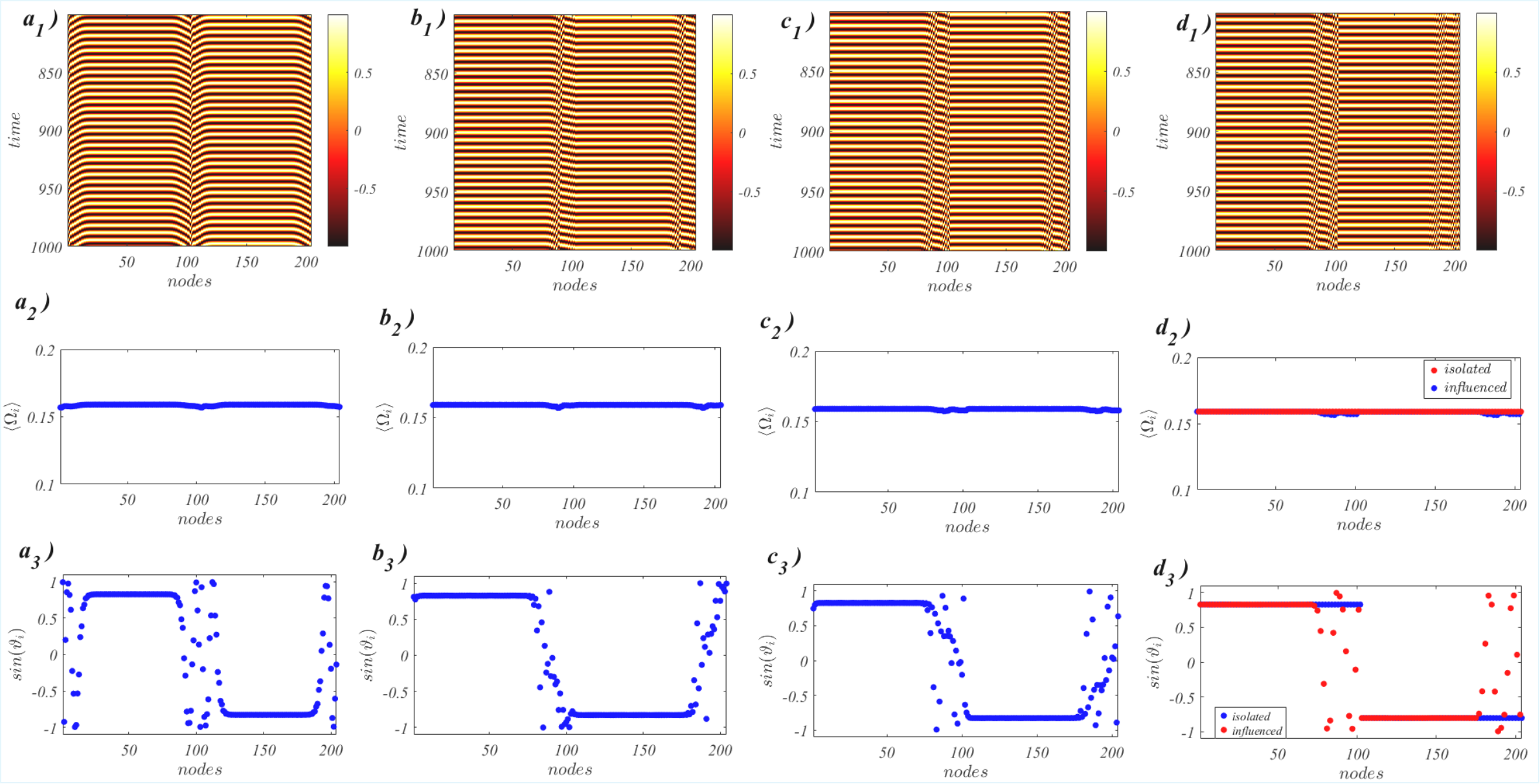}

		 \caption{Analysis of the dynamics of the phase reduced model on a $1$-directed $2$-hyperring of $204$ nodes. The first row shows the spatiotemporal diagrams {for the $\sin$ of the phases $\vartheta_i$}, the second row the {average} frequencies, and the last row the phases. The directionality parameter $p$ is varied with the columns: 
(a1, a2, a3) phase chimera states for \( p = 1 \)  ;  
(b1, b2, b3) phase chimera states for \( p = 0.2 \) ;  
(c1, c2, c3) phase chimera states for \( p = 0.1 \) ;  
(d1, d2, d3) phase chimera states for \( p = 0 \).  The coupling strength is \( \epsilon = 0.015 \) and the (angular) frequency is $\omega=1$.  }

\label{phase_reduction_HO}

\end{figure*}

  \begin{figure*} [ht!]
			\includegraphics[width=0.98\textwidth]{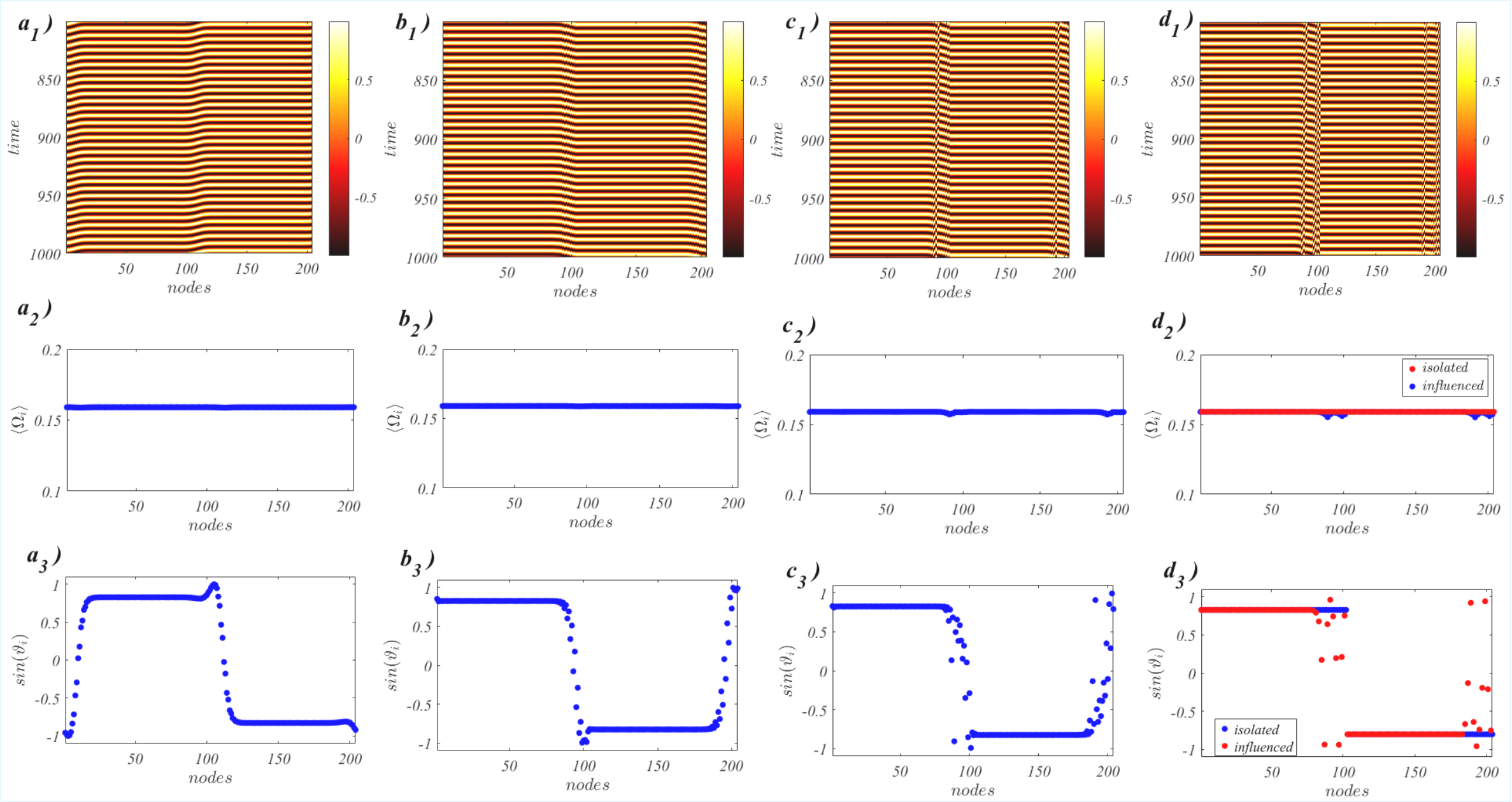}

		 \caption{Analysis of the dynamics of the phase reduced model on a clique-projected network of $204$ nodes. The first row shows the spatiotemporal diagrams {for the $\sin$ of the phases $\vartheta_i$}, the second row the {average} frequencies, and the last row the phases. The directionality parameter $p$ is varied with the columns: 
(a1, a2, a3) coherent clusters for \( p = 1 \) ;  
(b1, b2, b3) coherent clusters for \( p = 0.2 \) ;  
(c1, c2, c3) phase chimera states for \( p = 0.1 \) ;  
(d1, d2, d3) phase chimera states for \( p = 0 \).  The coupling strength is \( \epsilon = 0.015 \) and the (angular) frequency is $\omega=1$.  }
    \label{phase_reduction_cpn}

\end{figure*}

The patterns observed in the previous {Section~\ref{sec:4}} are phase chimeras and are characterized by (almost) constant amplitudes. Hence, they should be also observed in a ``pure'' phase model, i.e., where amplitudes do not play any role. To validate this claim, we apply the phase reduction method \cite{nakao2016phase,kuramoto2019concept,pietras2019network,monga2019phase} to the Stuart-Landau oscillators. The main idea behind the phase reduction method consists of reducing a given oscillatory system to a phase model, i.e., Kuramoto-type \cite{kuramoto1975,Kuramoto,acebron2005kuramoto}. {In a nutshell (see \cite{nakao2016phase,monga2019phase,pietras2019network} for a tutorial and more detailed explanations)}, starting from a system of identical\footnote{Note that the general theory works also for the case of non-identical oscillators, as long as the differences are small, namely, the frequency of each oscillator $i$ is such that $\omega_i=\omega+\delta\omega_i$, with $\delta\omega_i<<1$.} weakly coupled self-sustained oscillators, e.g., 
\begin{equation}
    \dot{\vec{x}}_i=\vec{f}(\vec{x}_i)+\epsilon\sum_{j=1}^N A_{ij}\vec{g}_{ij}(\vec{x}_j,\vec{x}_i),
\end{equation} 
the system can be reduced to the sole phase equations, i.e., \begin{equation}
        \dot{\vartheta}_i=\omega+\epsilon\sum_{j=1}^N A_{ij}\vec{Z}(\vartheta_i)\cdot\vec{g}_{ij}(\vartheta_j,\vartheta_i),
\end{equation} 
where $\omega$ is the frequency of the $i$-th oscillator and $\vec{Z}$ is the phase sensitivity function, whose expression can be obtained analytically only for Stuart-Landau \cite{nakao2016phase} and weakly nonlinear oscillators \cite{leon2023analytical}. {Note that the phase reduction is an approximation and further expansions need to be performed to obtain a more accurate description \cite{leon2019phase,bick2024higher}. Nonetheless, as long as the coupling strength is small and the amplitude does not play a relevant role in the dynamics (which is the case of phase chimeras described in the previous {Section~\ref{sec:4}}), the phase model obtained through phase reduction provides a very good approximation.} The phase reduction theory has been recently applied to systems with higher-order interactions \cite{ashwin2016hopf,leon2019phase,leon22a,Mau23,muolo2025pinning,leon2025theory}, allowing to obtain higher-order versions of the Kuramoto model\footnote{Note that there are several versions of the higher-order Kuramoto model not obtained through phase reductions, e.g., \cite{tanaka2011multistable,skardal2019abrupt,skardal2020higher,lucas2020multiorder}}.

By using the phase sensitivity function for the SL oscillator, i.e., \(\vec{Z}(\vartheta) = (-\sin\vartheta, \cos\vartheta)^\top\) and by expressing the vector field in polar coordinates remembering that, $\alpha$ being $1$, the amplitude of the limit cycle is $1$, \(\vec{X}_i = (\cos\vartheta_i, \sin\vartheta_i)^\top\), we can apply the phase reduction method, by obtaining the following equation for the phases
\begin{equation}
\frac{d\vartheta_i}{dt} = \omega + \epsilon \sum_{j_1, j_2} 
A_{i, j_1, j_2} \Phi(\vartheta_i, \vartheta_{j_1}, \vartheta_{j_2})\, ,
\end{equation}
where \(\omega\) is the natural frequency and \(\Phi\) is a coupling function dependent on phase differences. Note that, because of our choice of parameters in the Stuart-Landau, the variable $\omega$ corresponds to the frequency of both models. By applying the averaging method over one oscillation period \cite{averaging_sanders}, for the case of $2$-hyperring with cubic coupling, the coupling function \(\Phi\) can be explicitly computed as

\begin{widetext}
\begin{equation}
\Phi(\vartheta_i, \vartheta_{j_1}, \vartheta_{j_2})= 
 \frac{1}{4} \cos(\vartheta_{j_2} - \vartheta_i)
  + \frac{1}{4} \sin(\vartheta_{j_2} - \vartheta_i) 
 + \frac{1}{8} \cos(2\vartheta_{j_1} - \vartheta_{j_2} - \vartheta_i) 
 + \frac{1}{8} \sin(2\vartheta_{j_1} - \vartheta_{j_2} - \vartheta_i)
  - \frac{3}{8}.
\end{equation}
\end{widetext}

Let us observe that both pairwise and higher-order interactions are present in the phase model.

The same procedure can be repeated for the pairwise system, by obtaining
\begin{equation}
\dot{\vartheta}_i = \omega + \epsilon \sum_{j=1}^N A_{ij} \psi(\vartheta_i, \vartheta_j),
\end{equation}
where the coupling function \(\psi\) dependent on phase differences. Again, through averaging, we obtain the equations for the evolution of the phases

\begin{equation}
\dot{\vartheta}_i = \omega + \epsilon \sum_{j=1}^N A_{ij} \frac{3}{8} 
\left[ \cos(\vartheta_j - \vartheta_i) + \sin(\vartheta_j - \vartheta_i) - 1 \right],
\end{equation}
which is the well-known Kuramoto-Sakaguchi model \cite{sakaguchi1986soluble}.

We can now repeat the numerical analysis for the reduced models. The results presented in Fig.~\ref{phase_reduction_HO} refer to the dynamics on the $1$-directed $2$-hyperring, while Fig.~\ref{phase_reduction_cpn} shows the case of the clique-projected network. In both cases, the phase model behaves as the non-reduced model, confirming that the observed patterns are indeed phase chimera states.

\subsection{{Total variations and phase diagram}}
{To obtain a global view of the parameters space and the associated dynamical behaviors, we performed a dedicated series of numerical experiments for the hyperring composed by $N=204$ nodes, fixing all the parameters but $\epsilon$ and $p$, by computing the normalized total variations for phase, amplitude, and frequency as a function of those free parameters. The results are reported in Fig.~\ref{phase_diagramme}: panel (a) shows the normalized total phase variation, panel (b) the normalized total { angular frequency} variation, and panel (c) the normalized total amplitude variation, by using a color code. More in detail, the white regions correspond to vanishing total variations, while passing from blue to green the values of total variations increase. By gathering information from the three panels, four parameter regions, associated each with a precise type of chimera state, can be  identified. First, phase chimeras are observed when $V(\langle a \rangle) \sim 0$ and $V({\langle \omega\rangle}) \sim 0$, while the normalized total variation for the phase is large. This typically occurs for weak coupling strength, i.e., $\epsilon \lesssim 0.03$, across almost the entire range of $p$. A generic point in this region is represented by the black circle.  The next area of the parameter space returns amplitude-mediated chimeras, characterized by simultaneous large variations in { angular frequency}, amplitude and phase. The parameter region for which this occurs is roughly $0.03 < \epsilon \leq 0.3$ and $p \lesssim 0.5 $, and a typical example is represented by the black triangle. When the phase variation is low but not zero, and both {angular frequency} and amplitude variations vanish (white region in panels (b) and (c))), the system settles into a coherent state (see black diamond for a generic case). Finally, the ``vertical'' region associated to small $p$ and $\epsilon \gtrsim 0.03$, visible on the panel (a) (greenish), corresponds to large total phase variation and non-negligible total variation for amplitude and {angular frequency}. This can be classified as incoherent states.} {As explained when introducing the normalized total variation, let us remark that this scalar measure alone cannot determine the presence or not of a chimera state, and complementary approaches, such as direct visualization of the dynamics are required to establish the emergence of chimera states.}

\begin{figure*} [ht!]
			\includegraphics[width=1\textwidth]{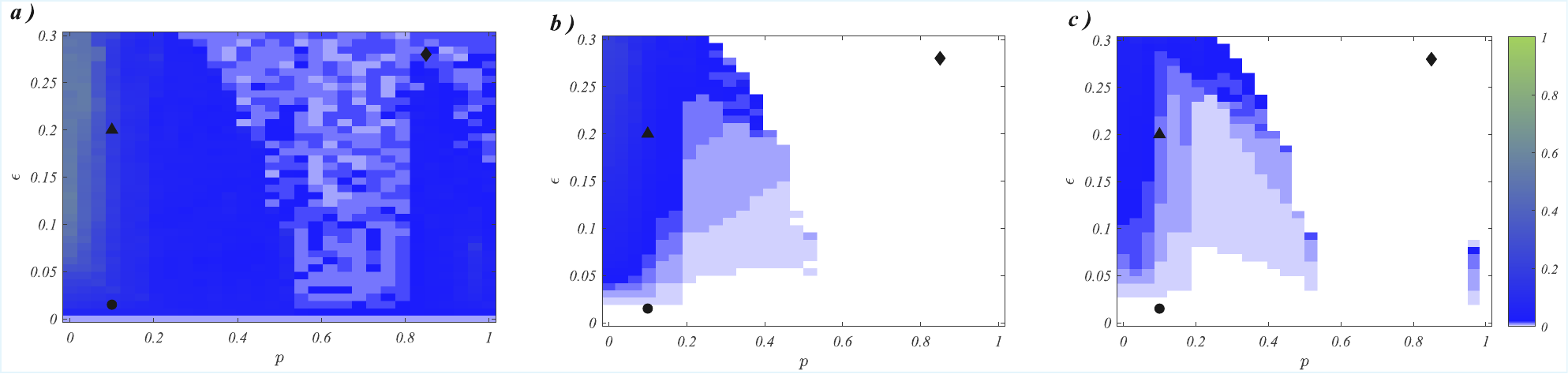} 
\caption{{Normalized total variation of phase (a), {angular frequency} (b), and  amplitude (c) as a function of $\epsilon$ and $p$ for a hyperring of $N=204$ nodes. The metrics are computed by numerically integrating the dynamical system on the time interval $[0,1000]$, and then by performing Fourier analysis on the resulting signals restricted to the window $[{900},1000]$. The values of the total variations are reported by using a color code: white regions indicate parameter ranges where the total variation is {quite small} or almost zero. {Light blue corresponds to a moderate level of variation, while dark blue corresponds to a relatively large variation, and, finally, very large values are depicted in green}. We emphasize {three} generic couples $(\epsilon,p)$ associated to specific {pattern} (see the text for a description of the different chimera states one can associate by gathering the information from the three panels). The black diamond indicates an example of $(\epsilon,p)$ with {quite} small phase variations or  vanishing variations in { angular frequency} and amplitude (coherent states), the black triangle one with important variations in amplitude, phase, and { angular frequency} (amplitude-mediated chimeras),  and the black circle shows one with significant phase variations but vanishing amplitude and { angular frequency} variations (phase chimeras).}  }
\label{phase_diagramme}
\end{figure*}

{
An analogous behavior can be observed in the case of the clique projection, as shown in Fig.~\ref{phase_diagramme_clique}, where, however, chimera states emerge for a smaller region of the parameters. In fact, a first conclusion can be directly drawn by looking at the three panels: for a larger range of parameter values, $\epsilon$ and $p$, the normalized total variations for phase, amplitude, and { angular frequency} reach lower values with respect to the hyperring case (lighter blue and large white regions). Equivalently stated, the higher-order support enhances the presence of chimera states measured by large values of the normalized total variations (the dark blue regions occupy a larger portion of the parameter space). Let us notice that, even if on a smaller scale, directionality induces chimera states also in the clique-projected network. Indeed, one can observe that, for small but positive values of $p$, one can have phase chimeras if $\epsilon$ is small (e.g., below $\sim {0.03}$, as depicted by the dark blue region in {panel (a), and white regions in the remaining panels}) and {amplitude-mediated chimeras} for the values of {$\epsilon > 0.03$} (dark blue region in panel (a), and light blue regions in the remaining panels). Larger values of $p$ diminish the presence of chimera states and return coherent ones. Finally, completely asymmetric networks, i.e., $p\sim 0$, return incoherent states (greenish vertical region on panel (a)).} {Once again, we remark that a complete comparison of the higher-order and pairwise setting is not possible, due to the different nature of the couplings. However, even by looking at the sole directionality parameter $p$, it is easy to observe a striking difference between the two settings.}
\begin{figure*} [ht!]
			\includegraphics[width=1\textwidth]{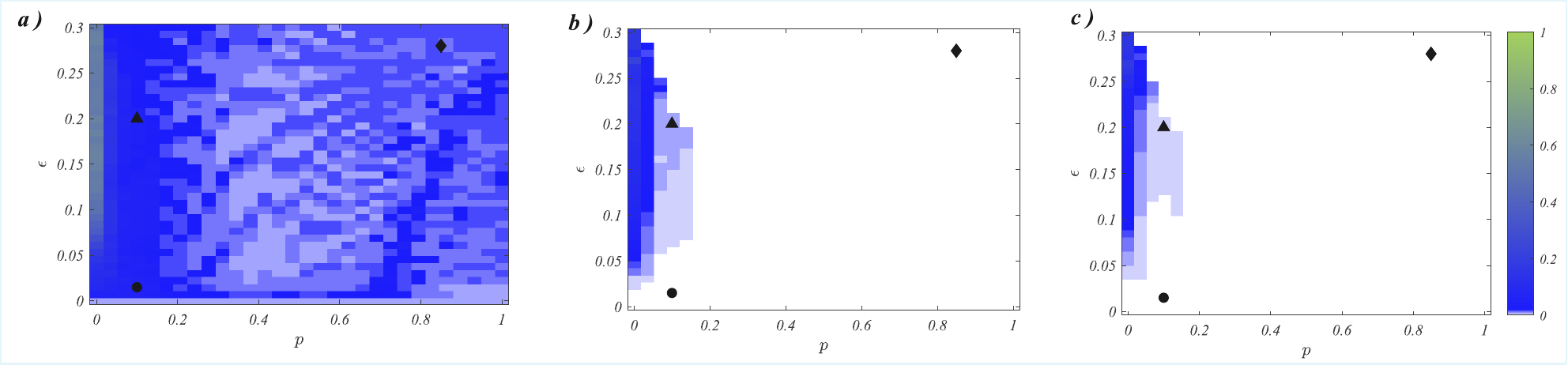} 
\caption{{Normalized total variation of phase (a), {angular frequency}  (b), and  amplitude (c) as a function of $\epsilon$ and $p$ for a clique-projected network of $N=204$ nodes. The normalized total variations are computed by numerically integrating the dynamical system on the time interval $[0,1000]$, and then by performing Fourier analysis on the resulting signals restricted to the window $[{900},1000]$. The values of the total variations are reported by using the same color code of Fig.~\ref{phase_diagramme}: white region indicates parameters ranges where the total variation {quite small} or almost zero.  {Light blue corresponds to a moderate level of variation, while dark blue corresponds to a relatively large variation and, finally, very large values are depicted in green}. We also emphasize {three} generic couples $(\epsilon,p)$ associated to specific {pattern} (see the text for a description of the different chimera states one can associate by gathering the information from the three panels). The black diamond indicates an example of $(\epsilon,p)$ with small phase variations {or}  vanishing variations in { angular frequency} and amplitude (coherent states), the black triangle one with important variations in amplitude, phase, and { angular frequency} (amplitude-mediated chimeras), and the black circle shows one with significant phase variations but vanishing amplitude and { angular frequency} variations (phase chimeras).}}
\label{phase_diagramme_clique}
\end{figure*}

{
The analysis performed so far has considered a regular hyperring with fixed size, $N=204$ nodes. However, it can be interesting to study the impact of the system size on its dynamics. For this reason, we computed the normalized total variations for hyperrings of increasing sizes, i.e., for $N \in \{10,\dots,408\}$, and two choices of the coupling strength ($\epsilon=0.2$ and $\epsilon=0.015$), while fixing all remaining parameters. Fig.~\ref{variation_in_terms_of_N_AMC} shows the dependence of the total variations of amplitude (panel (a)), { angular frequency} (panel (b)), and phase (panel (c)) as a function of the number of nodes in the case $\epsilon=0.2$ and $p=0.1$. A clear trend emerges from an eyeball analysis: the normalized total variations decrease once the system size increases. This is not surprising, as the chimera states are localized at the border between the two clusters of the initial conditions. In fact, the total variation remains large enough for the three considered variables, to indicate that a progressive increase in the number of nodes always preserves the presence of amplitude-mediated chimeras. Similarly, in Fig.~\ref{variation_in_terms_of_N_PC} we report analogous results but for a smaller value of the coupling strength, $\epsilon=0.015$, and still $p=0.1$, again for a number of oscillators $N \in \{10,\dots,408\}$. In this case, both the normalized total variation of amplitude and { angular frequency} vanish, while the normalized total phase variation shows an initial non-monotone trend to eventually steadily decrease, for the same reason as before. This demonstrates that by increasing $N$, the system preserves the presence of phase chimera. Note that for small values of $N$ it is difficult to detect a clear chimera state, as was shown in \cite{muolo2024phase}.}
\begin{figure*} [ht!]
			\includegraphics[width=1\textwidth]{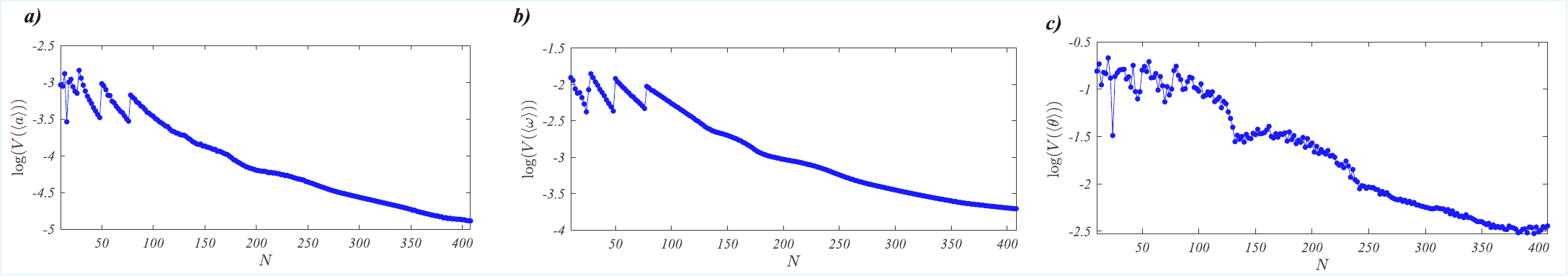} 
\caption{ {{Logarithm of the }normalized total variation of amplitude (a), {angular frequency} (b), and phase (c) as a function of the number of oscillators $N\in\{10,\dots, 408\}$ for the hyperring. %with $N \in \{\, n \in \mathbb{N} \mid 10 \leq n \leq 408,\; n \equiv 0 \pmod{2} \,\}$. 
The equations of motion are computed over the time interval $[0,1000]$, and then the Fourier analysis is performed on the window $[{900},1000]$. The coupling strength has been set to $\epsilon = 0.2$ and $p= 0.1$, values for which to amplitude-mediate chimeras emerge.}}
\label{variation_in_terms_of_N_AMC}
\end{figure*}

\begin{figure*} [ht!]
			\includegraphics[width=1\textwidth]{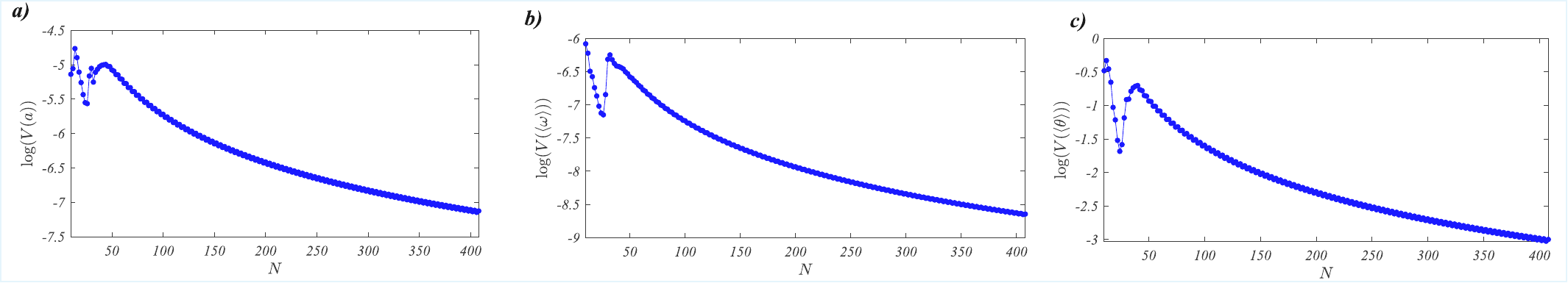} 
\caption{ {{Logarithm of the }normalized total variation of amplitude (a), {angular frequency} (b), and phase (c) as a function of the number of oscillators $N\in\{10,\dots,408\}$ for the hyperring.
%with $N \in \{\, n \in \mathbb{N} \mid 10 \leq n \leq 408,\; n \equiv 0 \pmod{2} \,\}$. 
The dynamics are computed over the time interval $[0,1000]$, and then Fourier analysis is performed on the window $[{900},1000]$. The coupling strength is $\epsilon = 0.015$ and $p=0.1$, setting for which phase chimeras emerge.}}
\label{variation_in_terms_of_N_PC}
\end{figure*}

%%%%%%%%%%%%%%%%%%%%%%%%%%%%%%%%%%%%%%%%%%%%%%%%%%%%%%%%%%%%%%%%%%%%

\section{Conclusion}

In this work, we have studied the emergence of chimera states in systems of oscillators coupled via directed {higher-order interactions, namely, $m$-directed hypergraphs}. {Our numerical study showed the presence of amplitude-mediated chimeras, which seem to be induced by the directionality, especially in the higher-order setting. Moreover, the combination of directionality and higher-order lead to the emergence of traveling amplitude-mediated chimeras, while analogous patterns are stationary in the pairwise setting, at least for the range of couplings explored.} Moreover, we have shown that phase chimera states observed on symmetric hypergraphs are preserved when directionality is induced. {These synchronization patterns are greatly enhanced by the presence of higher-order interactions.} The nature of phase chimeras was further validated through phase reduction theory, a perturbative technique allowing to describe any limit cycle oscillator through a phase equation. The phase model obtained in this way showed an analogous chimera pattern, confirming that what has been observed is indeed a phase chimera.
{Further analysis of this phenomenon could make use of linear stability analysis techniques, such as the one proposed in~\cite{bayani2024transition}; however, this analysis remains challenging because those methods rely on the Master Stability Function~\cite{fujisaka1983stability,pecora1998master}{, allowing us} to infer about the local stability of the homogeneous synchronous solution, while chimera states are mostly observed from inhomogeneous initial conditions, as is the case of this work. {For the case of the 2-directed 2-hyperedge (see appendix \ref{appB}), {an increasing directionality seems to impede the onset of chimera states.} A further study could focus on a deeper analysis to understand how these $m$-directed structures affect symmetry-breaking in coupled oscillators.}} {Further studies could also investigate the present of confounding factor and a different symmetrization method to vary directionality could be used, as done in~\cite{gallo2022synchronization}.} 

{Lastly, let us further remark that an exact matching of the coupling between the higher-order and pairwise case is not possible, due to their different nature. However, the comparison provided in this work may turn useful in light of possible applications, where the coupling strength could be associated to a cost or an energy measure. In fact, despite having dealt with} a toy model{, real-world} interactions, especially those in the brain, are in general non-reciprocal and higher-order, this work makes a step forward towards more realistic settings in which chimera states can emerge.

\section*{Acknowledgements}
R.T.D. acknowledges Thierry Njougouo and Patrick Louodop for discussions. H.N. acknowledges JSPS KAKENHI 25H01468, 25K03081, and 22H00516 for financial support. R.M. acknowledges JSPS KAKENHI 24KF0211 for financial support.

\section*{Author's contribution} 
R.T.D.: conceptualization, software, methodology, investigation, visualization, formal analysis, validation, writing -- review and editing. T.C.: software, methodology, visualization, supervision, writing -- review and editing. H.N.: methodology, writing -- review and editing.  R.M.: conceptualization, methodology, visualization, supervision, writing -- original draft, writing -- review and editing. All authors read and approved the manuscript.

%%%%%%%%%%%%%%%%%%%%%%%%%%%%%%%%%%%%%%%%%%%%%%%%%%%%%%%%%%%%%%%%%%%%%%%%%%%%%%%%%%%%%%%%%%%%%%%%%%%%%%%%%%%%%%%%%%%%%%%%%%%%%%%%%%%%%%%%%%%%%%%%%%%%%%

%

%%%%%%%%%%%%%%%%%%%%%%%%%%%%%%%%%%%%%%%%%%%%%%%%%%%%%%%%%%%%%%%%%%%%%%%%%%%%%%%%%%%%%%%%%%%%%%%%%%%%%%%%%%%%%%%%%%%%%%%%%%%%%%%%%%%%%%%%%%%%%%%%%%%%%%

\appendix
\onecolumngrid

\section{$1$-directed hypergraphs with different orientation}
\label{appA}
\setcounter{equation}{0}
\renewcommand{\theequation}{A\arabic{equation}}
\setcounter{figure}{0}
\renewcommand{\thefigure}{A\arabic{figure}}

In this section, we also consider a different orientation of the $1$-directed $2$-hyperring, namely, $q_2 = 1$, and $q_1 = q_3 = p$, so that the directionality is towards nodes that are not junction nodes. Also in this setting, we can obtain amplitude-mediated chimeras (not traveling), shown in Fig. \ref{f4}, and phase chimeras, shown in Fig. \ref{f5}. Such states are not observed for the corresponding clique-projected network (results not shown).

    \begin{figure*} [ht!]
			\includegraphics[width=0.98\textwidth]{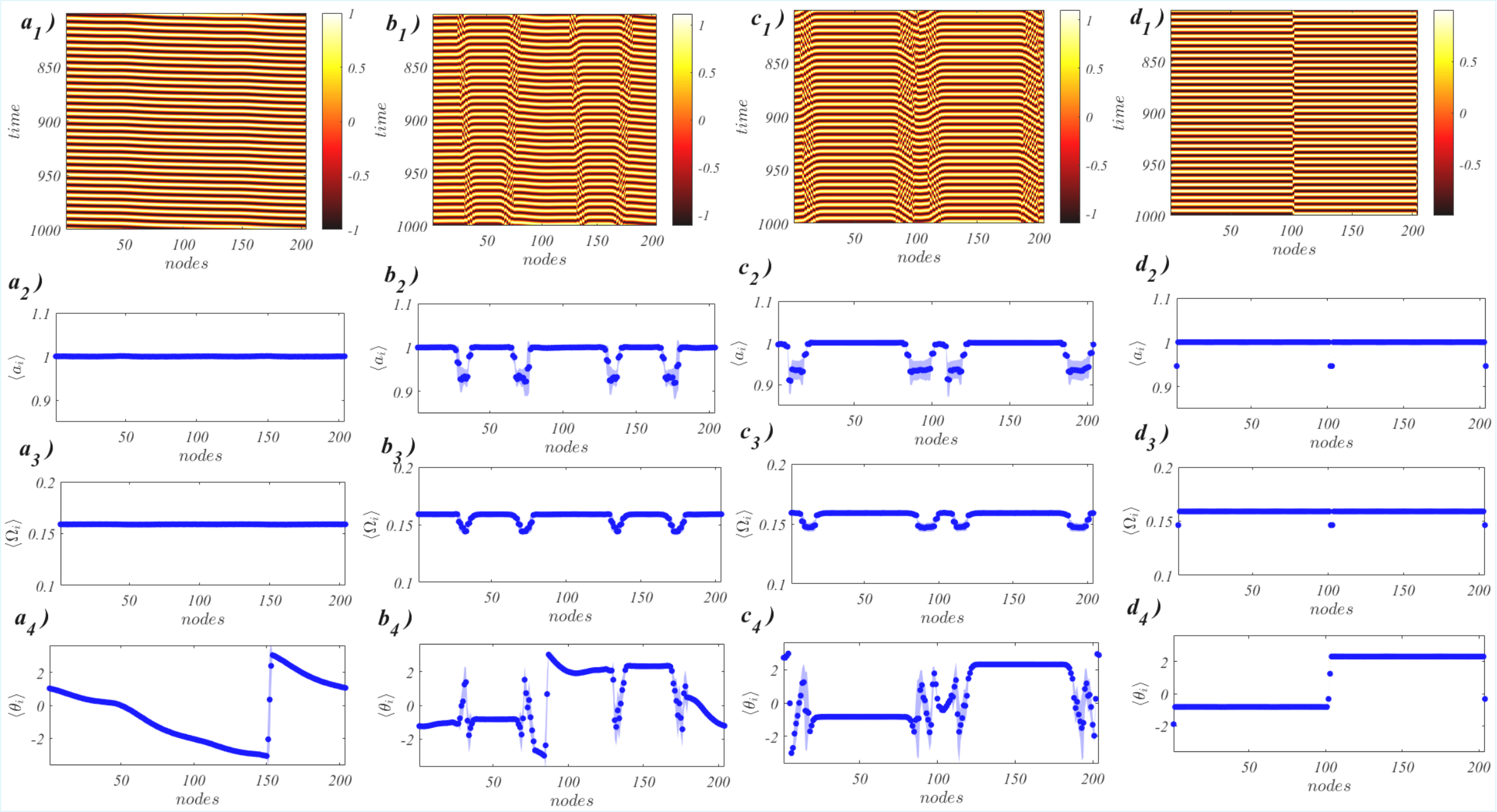}

		 \caption{Analysis of the dynamics on a $1$-directed $2$-hyperring of $204$ nodes with a different orientation with respect to the Main Text. The first row shows the spatiotemporal diagrams {for the $y$ variable (the behavior of the $x$ variable is analogous)}, the second row the {average} amplitudes, the third row the {average} frequencies, and the last row the {average} phases. The directionality parameter $p$ is varied with the columns: 
(a1, a2, a3, a4) coherent behavior for \( p = 1 \)  {(with $V(\langle a \rangle) \approx 2.45\,{\times}10^{-4} $, $V(\langle \omega \rangle) = 1.2\,{\times}10^{-4}$, and $V(\langle \theta \rangle) \approx 0.0192$)} ;  
(b1, b2, b3, b4) amplitude-mediated chimera state for \( p = 0.1 \) {(with $V(\langle a \rangle) \approx 0.0195$, $V(\langle \omega \rangle) = 0.016$, and $V(\langle \theta \rangle) \approx 0.0875$)} ;  
(c1, c2, c3, c4)  amplitude-mediated chimera states for \( p = 0.05\) {(with $V(\langle a \rangle) \approx 0.0323 $, $V(\langle \omega \rangle) = 0.0248$, and $V(\langle \theta \rangle) \approx 0.1036$)};  
(d1, d2, d3, d4) coherent clusters for \( p = 0 \)  {(with $V(\langle a \rangle) \approx 2.12\,{\times}10^{-3}$, $V(\langle \omega \rangle) = 0.0031$, and $V(\langle \theta \rangle) \approx 9.80\,{\times}10^{-3}$)}.  The model parameters are $\alpha=1$ and $\omega=1$, and the coupling strength is \( \epsilon = 0.2 \).   {The shaded light blue area represents the standard deviation of the quantity, computed over $10$ consecutive sub-intervals, and quantifies the temporal variability of the node dynamics around its mean value.} }
    \label{f4}

\end{figure*}

    \begin{figure*} [ht!]
			\includegraphics[width=0.98\textwidth]{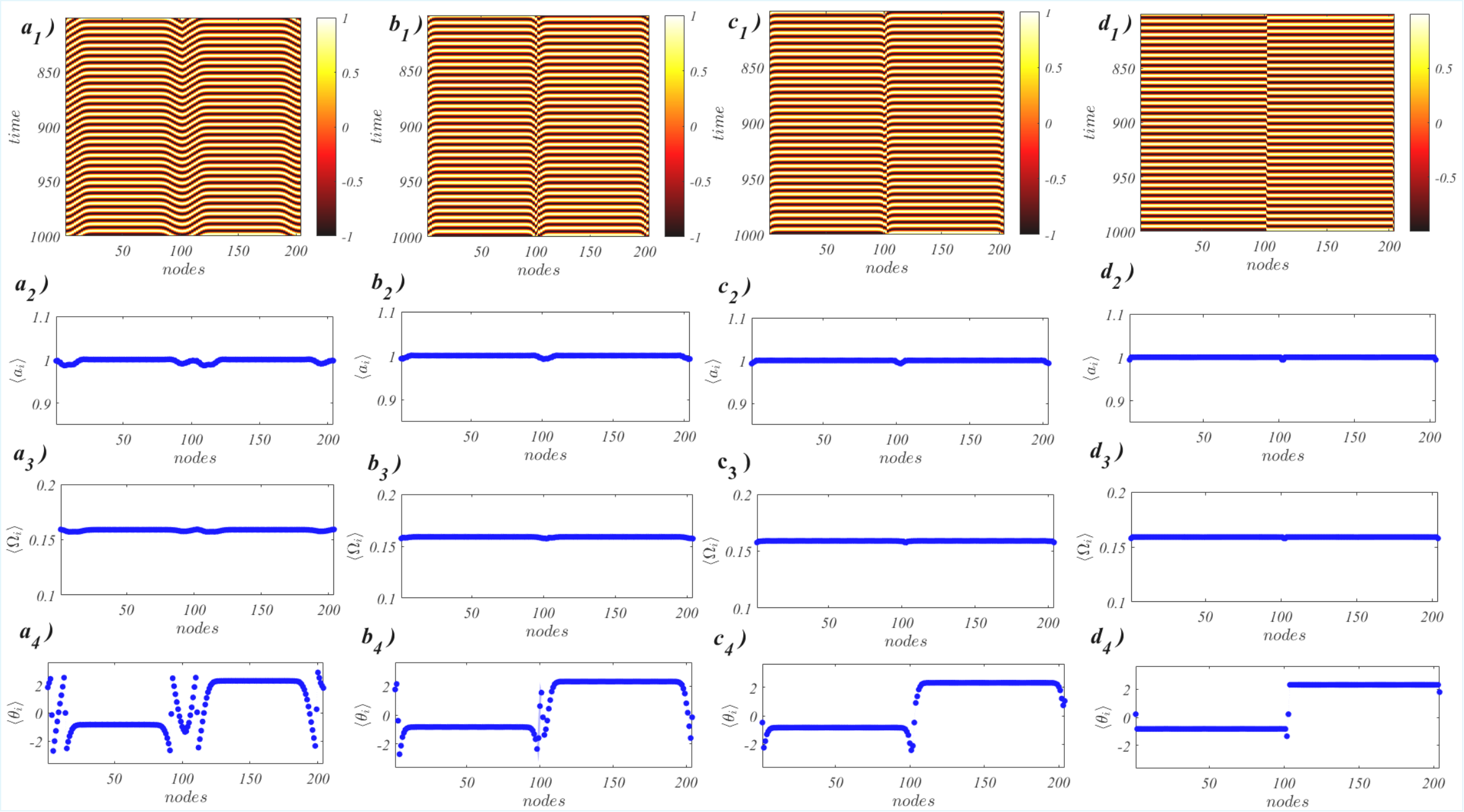}

		 \caption{Analysis of the dynamics on a {$1$}-directed $2$-hyperring of $204$ nodes with a different orientation with respect to the Main Text. The first row shows the spatiotemporal diagrams {for the $y$ variable (the behavior of the $x$ variable is analogous)}, the second row the {average} amplitudes, the third row the {average} frequencies, and the last row the {average} phases. The directionality parameter $p$ is varied with the columns: 
(a1, a2, a3, a4) phase chimera state for \( p = 1 \) {(with $V(\langle a \rangle) \approx 2.86\,{\times}10^{-3} $, $V(\langle \omega \rangle) = 5.0\,{\times}10^{-4}$, and $V(\langle \theta \rangle) \approx 0.0529$)};  
(b1, b2, b3, b4) phase chimera state for \( p = 0.2 \) {(with $V(\langle a \rangle) \approx 8.63\,{\times}10^{-4}$, $V(\langle \omega \rangle) = 4.40\,{\times}10^{-4}$, and $V(\langle \theta \rangle) \approx 0.0393$)};  
(c1, c2, c3, c4)  phase chimera state for \( p = 0.1 \) {(with $V(\langle a \rangle) \approx 4.94\,{\times}10^{-4} $, $V(\langle \omega \rangle) = 3.0\,{\times}10^{-4}$, and $V(\langle \theta \rangle) \approx 0.0199$)};  
(d1, d2, d3, d4) coherent clusters for \( p = 0 \) {(with $V(\langle a \rangle) \approx 2.24\,{\times}10^{-4} $, $V(\langle \omega \rangle) = 3.0\,{\times}10^{-4}$, and $V(\langle \theta \rangle) \approx 9.80\,{\times}10^{-3}$)}. The model parameters are $\alpha=1$ and $\omega=1$, and the coupling strength is \( \epsilon = 0.02 \).  {The shaded light blue area represents the standard deviation of the quantity under consideration, computed over $10$ adjacent sub-intervals, and quantifies the temporal variability of the node dynamics around its mean value.} }
    \label{f5}

\end{figure*}

\section{$2$-directed hypergraphs}
\label{appB}
\setcounter{equation}{0}
\renewcommand{\theequation}{B\arabic{equation}}
\setcounter{figure}{0}
\renewcommand{\thefigure}{B\arabic{figure}}

If we now consider $2$ nodes in the head of a directed hypergraph and $1$ node in the tail, we obtain a $2$-directed $2$-hyperring, as represented in Fig. \ref{hypergraphe_2_diridé_2_hyperring}. As for the case of $1$-directed, we can easily obtain its corresponding clique-projected network, shown in Fig. \ref{fig:clique_projection_2}. If we start from the same setting of Fig. \ref{f1121} where the symmetric case exhibits phase chimeras, we see that such pattern is not conserved when inducing the directionality, as shown in Fig.~\ref{f112}. In fact, already from $p\lesssim 0.9$ the phase chimera vanishes. In Fig. \ref{clique11}, we show the results for the clique-projected network, where no patterns are observed. Interestingly, there are no patterns even when the directionality is such that there are isolated nodes, at contrast with the case of the Main Text.

\begin{figure*}[ht!]
    \centering
    \includegraphics[width=0.9\textwidth]{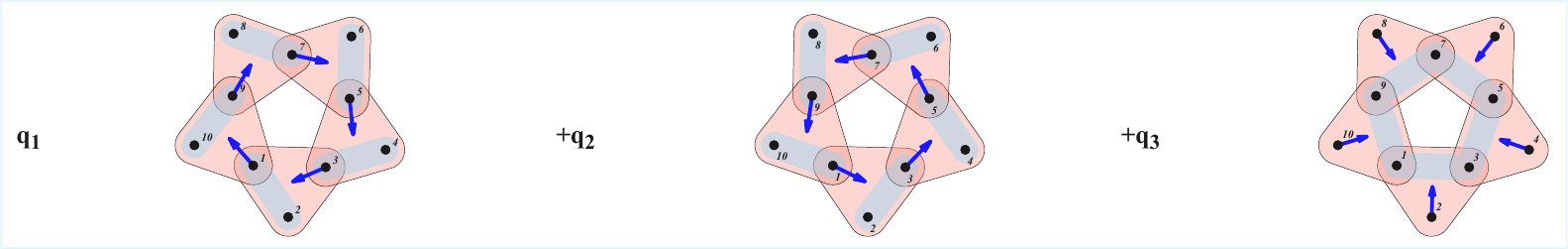}
\caption{Schematically representation of a family of \textbf{$2$-directed $2$-hyperrings}. In the graphical representation, the heads of the hyperedges are highlighted in blue, while arrows indicate the directionality of the connections between the nodes.}

    \label{hypergraphe_2_diridé_2_hyperring}
\end{figure*}

\begin{figure*}[ht!]
    \centering
    \includegraphics[width=0.9\textwidth]{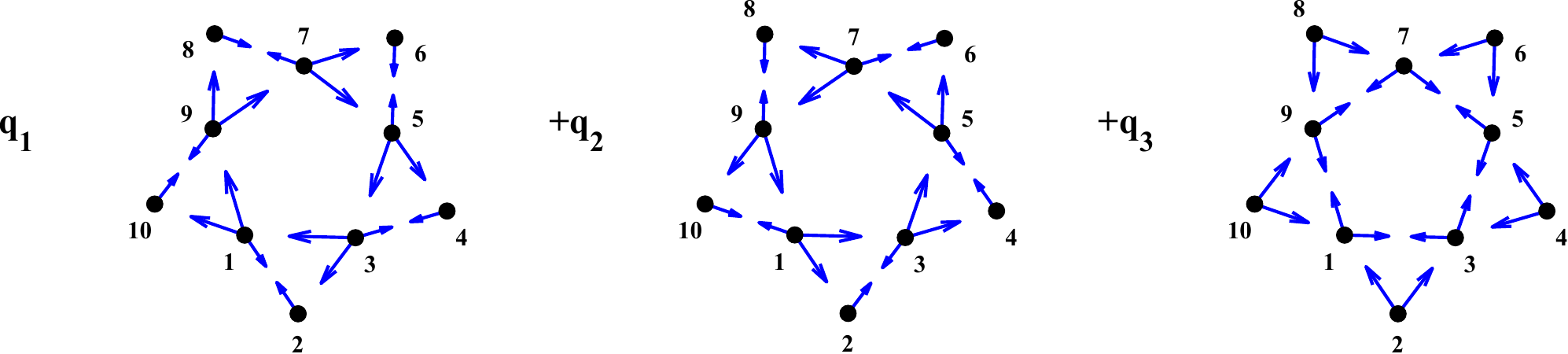}
    \caption{Schematic representation of a family of clique-projected networks, corresponding to the $2$-directed $2$-hyperring of the previous Figure. Note that in the case of 2-directed 2-hyperrings, the two head nodes interact with each other. }
    \label{fig:clique_projection_2}
\end{figure*}

 \begin{figure*} [ht!]
			\includegraphics[width=0.98\textwidth]{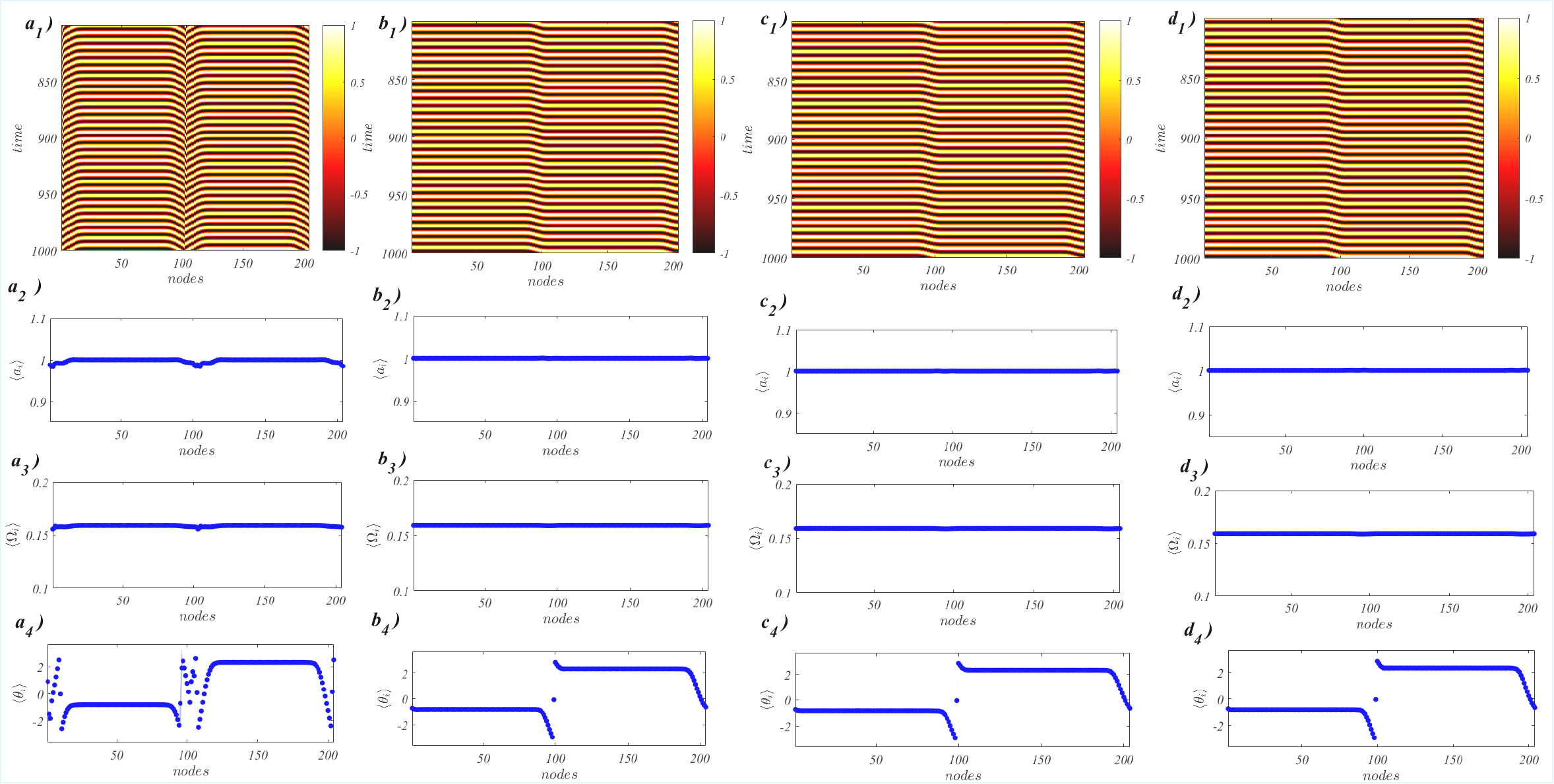} 
            		 \caption{Analysis of the dynamics on a $2$-directed $2$-hyperring of $204$ nodes. The first row shows the spatiotemporal diagrams {for the $y$ variable (the behavior of the $x$ variable is analogous)}, the second row the {average} amplitudes, the third row the {average} frequencies, and the last row the {average} phases. The directionality parameter $p$ is varied with the columns: 
(a1, a2, a3, a4) phase chimera state for \( p = 1 \) {(with $V(\langle a \rangle) \approx 1.91\,{\times}10^{-3} $, $V(\langle \omega \rangle) = 8.0\,{\times}10^{-4}$, and $V(\langle \theta \rangle) \approx 0.0659$)};  
(b1, b2, b3, b4) coherent clusters for \( p = 0.2 \) {(with $V(\langle a \rangle) \approx 5.6\,{\times}10^{-5}$, $V(\langle \omega \rangle) = 5.66 \, {\times}10^{-5}$, and $V(\langle \theta \rangle) \approx 9.92\,{\times}10^{-3}$)} ;  
(c1, c2, c3, c4)  coherent clusters for \( p = 0.1 \) {(with $V(\langle a \rangle) \approx 3.5\,{\times}10^{-4}$, $V(\langle \omega \rangle) = 5.02 \,{\times}10^{-5}$, and $V(\langle \theta \rangle) \approx 9.90\,{\times}10^{-3}$)} ;  
(d1, d2, d3, d4) coherent clusters for \( p = 0 \) {(with $V(\langle a \rangle) \approx 2.2\,{\times}10^{-5}$, $V(\langle \omega \rangle) = 5.02 \,{\times}10^{-5}$, and $V(\langle \theta \rangle) \approx 9.80\,{\times}10^{-3}$)} .  The model parameters are $\alpha=1$ and $\omega=1$, and the coupling strength is \( \epsilon = 0.015 \).  {The shaded light blue area represents the standard deviation of the quantity under scrutiny, computed over $10$ consecutive sub-intervals, and quantifies the temporal variability of the node dynamics around its mean value.}}
    \label{f112}
	\end{figure*}

%     \begin{figure*}[hbtp!]
%     \centering
%         \includegraphics[width=0.98\textwidth]{Fig15.pdf}

% \caption{Analysis of the dynamics on a clique-projected network of $204$ nodes. The first row shows the spatiotemporal diagrams, the second row the amplitudes, the third row the frequencies, and the last row the phases. The directionality parameter $p$ is varied with the columns: 
% (a1, a2, a3, a4) coherent clusters for \( p = 1 \) ;  
% (b1, b2, b3, b4) coherent clusters for \( p = 0.2 \) ;  
% (c1, c2, c3, c4)  coherent clusters for \( p = 0.1 \) ;  
% (d1, d2, d3, d4) coherent clusters for \( p = 0 \).  The model parameters are $\alpha=1$ and $\omega=1$, and the coupling strength is \( \epsilon = 0.015 \). }

%     \label{clique11}
% \end{figure*}

    \begin{figure*}[hbtp!]
    \centering
        \includegraphics[width=0.98\textwidth]{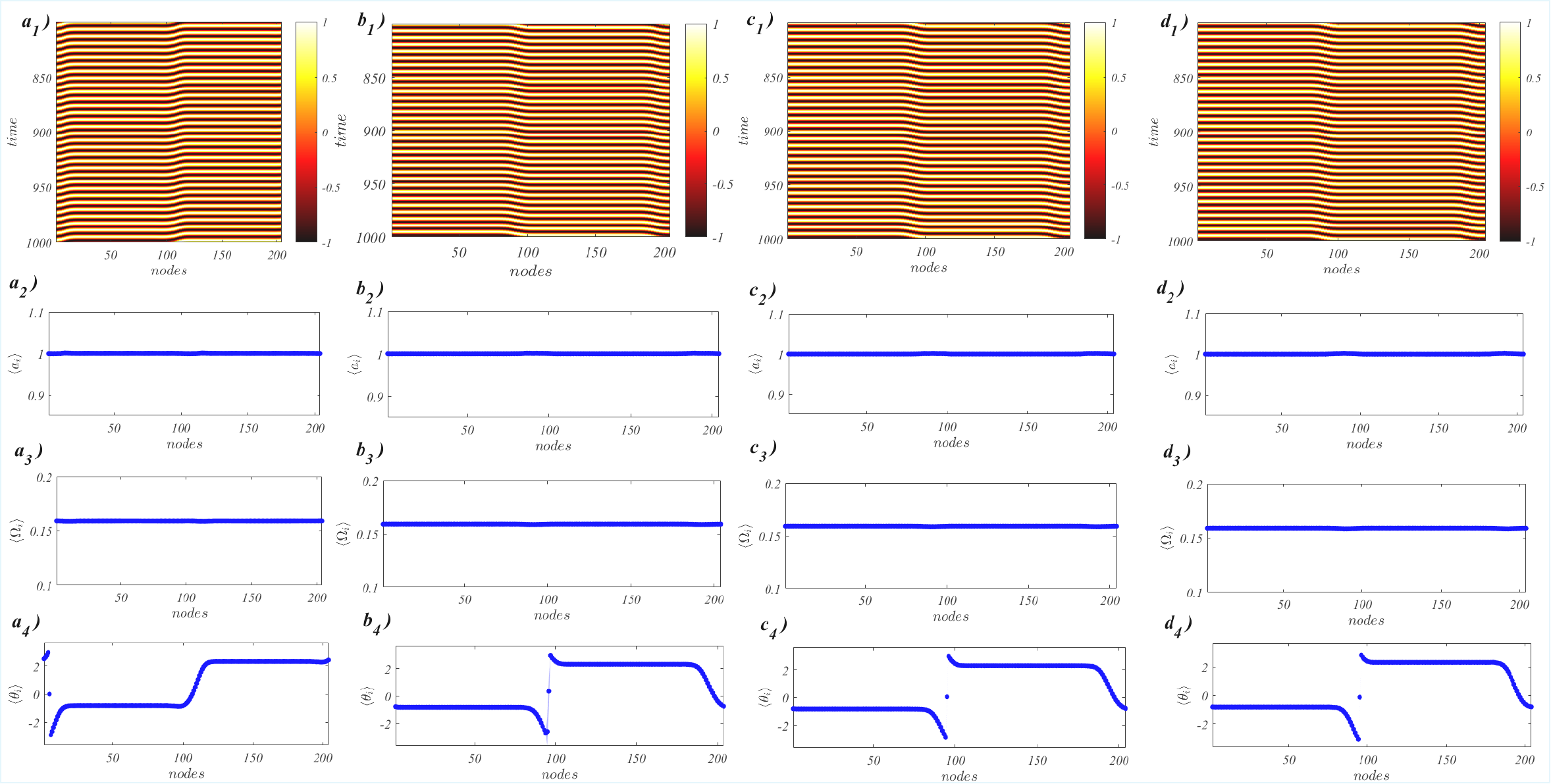}

\caption{Analysis of the dynamics on a clique-projected network of $204$ nodes. The first row shows the spatiotemporal diagrams {for the $y$ variable (the behavior of the $x$ variable is analogous)}, the second row the {average} amplitudes, the third row the {average} frequencies, and the last row the {average} phases. The directionality parameter $p$ is varied with the columns: 
(a1, a2, a3, a4) coherent clusters for \( p = 1 \) {(with $V(\langle a \rangle) \approx 2.24\,{\times}10^{-4}$, $V(\langle \omega \rangle) = 6.28 \,{\times}10^{-5}$, and $V(\langle \theta \rangle) \approx 0.0103$ )};  
(b1, b2, b3, b4) coherent clusters for \( p = 0.2 \) {(with $V(\langle a \rangle) \approx 8.3\,{\times}10^{-5}$, $V(\langle \omega \rangle) = 6.28 \, {\times}10^{-5}$, and $V(\langle \theta \rangle) \approx 0.0109$)} ;  
(c1, c2, c3, c4)  coherent clusters for \( p = 0.1 \) {(with $V(\langle a \rangle) \approx 6.5\,{\times}10^{-5}$, $V(\langle \omega \rangle) = 7.54 \, {\times}10^{-5}$, and $V(\langle \theta \rangle) \approx 9.81\,{\times}10^{-3}$)};  
(d1, d2, d3, d4) coherent clusters for \( p = 0 \) {(with $V(\langle a \rangle) \approx 4.3\,{\times}10^{-4}$, $V(\langle \omega \rangle) = 7.53 \, {\times}10^{-5}$, and $V(\langle \theta \rangle) \approx 9.85\,{\times}10^{-3}$)}. The model parameters are $\alpha=1$ and $\omega=1$, and the coupling strength is \( \epsilon = 0.015 \).   {The shaded light blue area represents the standard deviation of the quantity under study, computed over $10$ adjacent sub-intervals, and quantifies the temporal variability of the node dynamics around its mean value.} }

    \label{clique11}
\end{figure*}

\end{document}